\documentclass{aastex63}
\usepackage[utf8]{inputenc}
\usepackage[margin=0.9in]{geometry}
\usepackage{amsmath}
\usepackage{color}
\usepackage{indentfirst}
\usepackage{graphicx}
\graphicspath{ {./} }

\received{}
\revised{}
\accepted{}
\submitjournal{ApJ}

\shorttitle{Her X-1 35-day X-ray Cycle}
\shortauthors{Leahy \& Wang}
 
\begin{document}

\title{Swift/BAT and RXTE/ASM Observations of the 35-day X-ray Cycle of Hercules X-1}

\correspondingauthor{Denis Leahy}
\email{leahy@ucalgary.ca}

\author[0000-0002-0786-7307]{Denis Leahy}
\affiliation{Dept. of Physics $\&$ Astronomy, University of Calgary, University of Calgary,
Calgary, Alberta, Canada T2N 1N4}

\author{Yuyang Wang}
\affiliation{Dept. of Physics $\&$ Astronomy, University of Calgary, University of Calgary,
Calgary, Alberta, Canada T2N 1N4}

\begin{abstract}
Swift/BAT and RXTE/ASM observations have monitored the X-ray binary system Her X-1 for approximately
14.5 years each, and both were monitoring Her X-1 for a period of $\sim$5.5 years. 
Here we study the 35-day cycle using these observations. Using a cross-correlation method we
find the times of peaks of the 35-day cycles for $\sim$150 cycles observed by Swift/BAT and 
$\sim$150 cycles observed by RXTE/ASM. These cycles include $\sim$60 observed with both instruments.
The noise level of the RXTE/ASM measurements is larger than that of Swift/BAT, resulting in larger uncertainty in
peak times.
The distribution of 35-day cycle lengths can be fit with a Gaussian with mean 34.79 d and $\sigma$ of 1.1 d.
The distribution of orbital phases of 35-day cycle peaks is well fit by a uniform distribution, 
with 76\% of the cycles, plus a Gaussian distribution peaked at orbital phase $\sim 0.5$, with 24\% of the cycles.
We construct the long-term average 35-day lightcurve in the 15-50 keV
band from  Swift/BAT, and in the 2-12 keV band from RXTE/ASM.
The high energy band shows more variability in the Short High state and the low energy band shows more
variability in the Main High state. This is consistent with 
 a precessing
accretion disk model as cause of the 35-day cycle.
\end{abstract}


\section{Introduction}  \label{sec:intro}

Her X-1/HZ Her is an X-ray binary pulsar, that remains an object of current research. 
\citet{2019ApJ...871..152L} report spectral analysis of AstroSat Soft X-ray Telescope observations
of Her X-1 during  Low State, Turn-on, Main High State (hereafter MH) and MH during dip.
\citet{2016ApJ...831..194W} use the  NuSTAR X-ray spectrum of Her X-1 as a testbed for radiation-dominated radiative shock models of the accretion column.
 \citet{2015AJ....150....3S} compares long-term RXTE/ASM X-ray monitoring with AAVSO optical data
 to show the X-ray MH state fluence is considerably more variable than the optical fluence.
 \citet{2015ApJ...800...32L} utilized RXTE/PCA  X-ray eclipse observations to detect and measure the
 extended scattering corona in the binary system.
The masses and uncertainties of the neutron star (Her X-1,  $\sim$1.5 $M_\odot$) and its stellar companion (HZ Her, $\sim$2.3 $M_\odot$)  are reviewed in \citet{2014ApJ...793...79L} and \citet{1997MNRAS.288...43R}. 

The 35-day cycle, consisting of consecutive MH,  Low State, Short High State and Low State,  
 has been monitored with a number of satellites. 
 The rapid rise to MH occurs over a timescale of $\sim$3 hours (e.g. Fig. 3 of \citealt{2000ApJ...539..392S}) and is called turn-on.
23 cycles from RXTE/ASM observations were analysed by \citet{1998MNRAS.300..992S} and by \citet{1999ApJ...510..974S}.
These studies measured accurately the mean 35-day light curve for the first time, measured dip behaviour, and derived sets of turn-on times. 
\cite{2010ApJ...713..318L} and \citet{2013A&A...550A.110S} analyzed more than 10 years of RXTE/ASM observations to
characterize cycle length variability. \cite{2010ApJ...713..318L} analyzed turn-on phases of the 35-day cycle and found a uniform distribution with orbital phase.

Her X-1 occasionally shows lengthy time periods, Anomalous Low States or ALS, 
during which the 35-day cycle shuts off in X-rays, as first discovered by \citet{1985Natur.313..119P}, 
who suggested a change in disk structure as the cause.
ALS are further studied in  \cite{1999A&A...350L...5P} and \citet{2000ApJ...543..351C}, where
change of warp of the disk or decrease of disk inclination, respectively, were proposed as the causes of the ALS.
Subsequent work showed that the 35-day cycle continues in optical during X-ray anomalous low states \citep{2011MNRAS.418..437J}.
\cite{2010ApJ...715..897L} analyzed EUV observations of Her X-1 during anomalous low.
They modeled the orbital modulation of the EUV emission from the heated face HZ Her, 
and found a thicker accretion disk than for normal 35-day cycle. 
The thicker disk is sufficient to block the compact X-ray emission but not to block the emission from HZ Her.

Multi-wavelength studies of the Her X-1/HZ Her system are possible because different parts of the 
system radiate in optical, ultraviolet, EUV and X-rays.
Systematic variations of the 1.7 d optical lightcurve \citep{1976ApJ...209..562G}  give strong evidence 
for a Roche-lobe-filling precessing accretion disk. 
The physics of precessing accretion disks have been successfully modelled by \citet{1999MNRAS.308..207W},
 which yields a disk like that measured for Her X-1 by \citealt{2002MNRAS.334..847L}.
The EUV emission from Her X-1 is from the inner disk and from the irradiated surface of HZ Her (\citealt{1999ApJ...521..328L}, \citealt{2003MNRAS.342..446L}).
Accretion of matter onto the rotating neutron star produces X-rays ($>1$ keV).

The X-ray pulsations have been modelled as emission from an accretion column geometry subsequently modified by light-bending in the neutron star's gravity (\citealt{2004MNRAS.348..932L}, \citealt{2004ApJ...613..517L}).
The pulse shape changes systematically over the 35-day cycle (\citealt{2000ApJ...542..446L} and references therein).
The cause of the pulse shape changes is explained by systematic changes in obscuration by the precessing accretion
disk \citep{2000ApJ...542..446L}, or alternately by neutron star free precession (\citealt{2012int..workE..22P} and referemces therein).
\citet{2013A&A...550A.110S} compared turn-on times measured from the flux variations with time from RXTE/ASM observations
with turn-on measured using pulse profile variations mainly from RXTE/PCA observations. The two measurements of turn-on agree,
implying a single clock mechanism behind the 35-day flux cycle and the 35-day pulse shape change cycle and raising serious questions
about the free precession mechanism. Subsequently, \citet{2013MNRAS.435.1147P} devised a free precession model that gives
approximate agreement of the 35-day flux cycle with the 35-day pulse shape change cycle.

The atmosphere of HZ Her causes X-ray absorption which is detected during eclipse ingresses and egresses 
(\citealt{1988MNRAS.231...69D}, \citealt{1995MNRAS.276..607L}). 
 X-rays reflected off the companion star are detected during Low State and Short High State \citep{2015MNRAS.453.4222A}.
Timing of MH ingresses and egresses has enabled accurate determination of the radius of HZ Her \citep{2014ApJ...793...79L}. 
Overall, the regular time-variations in Her X-1 are understood as caused by the geometry of the
system, i.e. Roche-lobe filling companion, accretion disk and stream. 

The X-ray flux and pulse shapes both systematic changes over a 35-day cycle. 
These are caused by the precessing accretion disk in the system \citep{2000ApJ...539..392S}. 
The disk occults the line-of-sight to the neutron star during during  Low State, and to a lesser extent during MH Turn-on and Short High State.
The geometry of the accretion disk has been measured by applying tilted-twisted disk models 
(\citealt{2002MNRAS.334..847L}, \citealt{2004AN....325..205L} and \citealt{2020ApJ...889..131L}).
 The well-known X-ray absorption dips are caused by the accretion stream connecting HZ Her to the disk  (\citealt{2011MNRAS.418.2283I} and \citealt{2012MNRAS.425....8I}). 

Here we carry out a new study of the long term properties of the 35-day cycle of Her X-1,
by analysing Swift/BAT and RXTE/ASM monitoring observations.
In sections \ref{sec:obs} and  \ref{sec:analysis} we describe the observations and data analysis of the 35-day cycle lightcurve of Her X-1. 
In section \ref{sec:discussion}, we discuss the results and what they imply for the physical mechanisms for the 35-day cycle. 
We summarize the work in section \ref{sec:conclusion}.

\section{Observations\label{sec:obs}} 

\subsection{Swift/BAT Observations}

\begin{figure}[ht!]
  \includegraphics[width=1.0\linewidth]{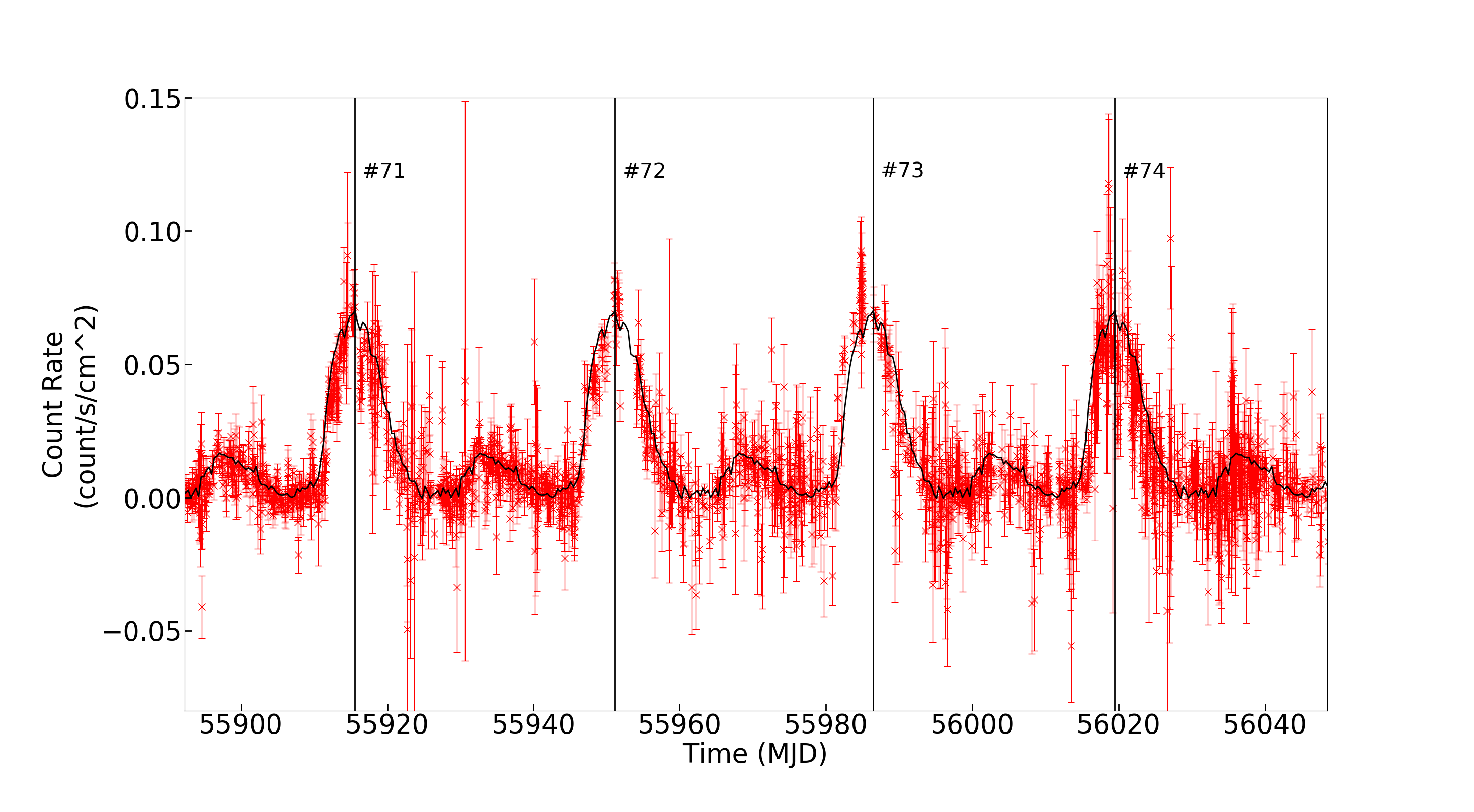}
   \includegraphics[width=1.0\linewidth]{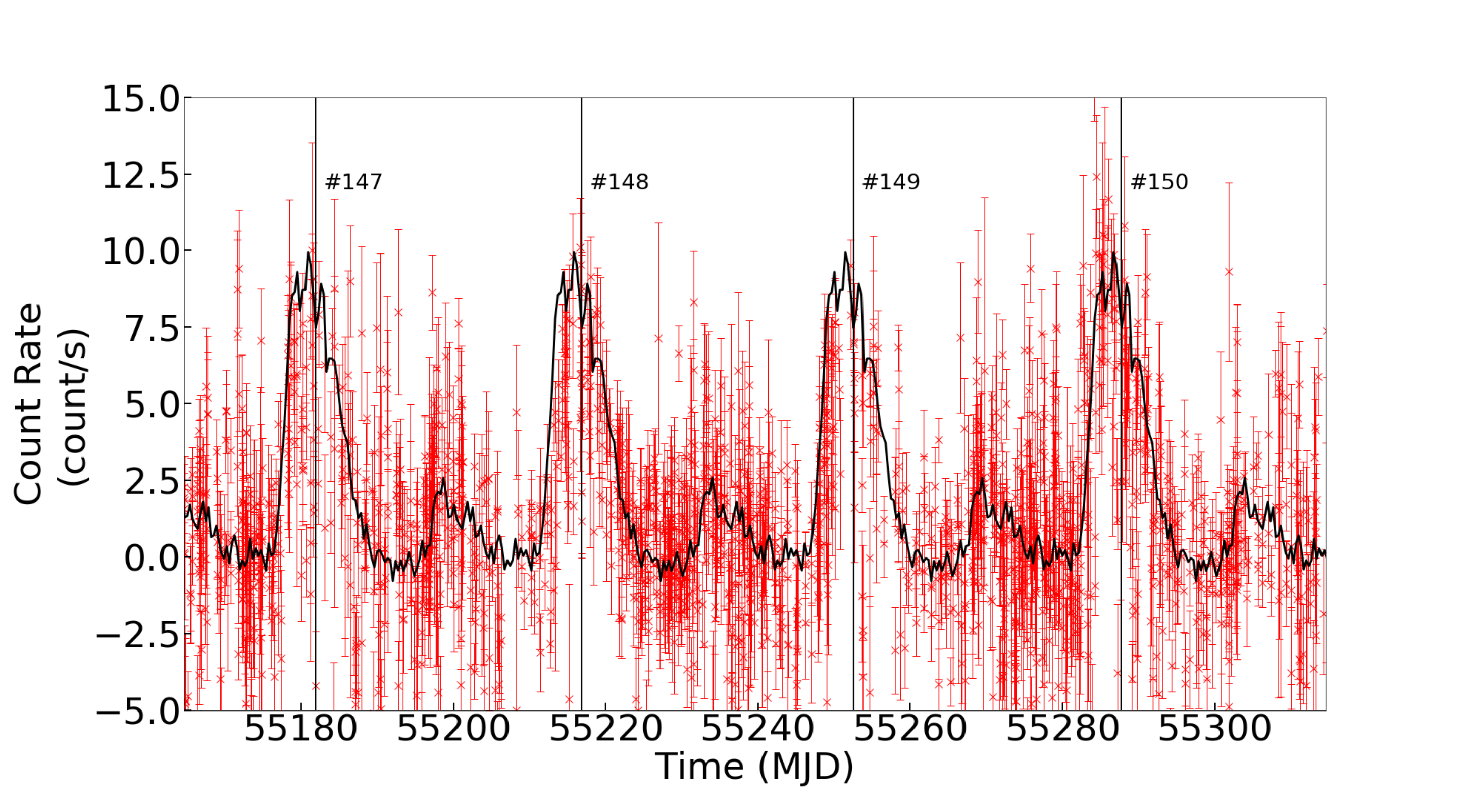}
    \caption{Top: Sample of the Swift/BAT lightcurve of Her X-1 between MJD55890 and MJD56050 (red points with error bars).
     The black line is the template lightcurve from the cross-correlation (CC) analysis and vertical black lines are times of peak of the CC function (see text for explanation).
    Bottom: Sample of the RXTE/ASM lightcurve of Her X-1 between MJD55160 and MJD55320 (red points with error bars). 
    The black line is the template lightcurve from the CC analysis and vertical black lines are times of peak of the CC function.
      \label{fig:BATdata}}
\end{figure}

The Burst Alert Telescope (BAT) onboard the
 Neil Gerels Swift Observatory 
  is described by \citet{2013ApJS..209...14K}.
BAT is used to monitor X-ray sources on a regular basis and to create long-term lightcurves for these sources.  
The data is available at the website https://swift.gsfc.nasa.gov/results/transients/.
For Hercules X-1, we downloaded the data for the energy range of 15-50 keV, covering 14 years of observation
with 71165 points from February 2005 to November 2019.
 The top panel of Figure~\ref{fig:BATdata} shows a sample of the Swift/BAT lightcurve of Her X-1.
The 35-day cycle of Her X-1, as quantified by e.g. \citet{1999ApJ...510..974S} and \citet{2004AN....325..205L},
is clearly seen in the data, with the large peak for MH state and the smaller peak for short high state.

The Swift/BAT orbital lightcurve has time as mission time in unit of seconds. 
We used the standard conversion into fractional Modified Julian Date in Terrestrial Time: 
$time_{MJD} = \frac{time_{BAT}}{24\times3600} + MJD_{ref}$
with $MJD_{ref} = 51910.00074287038$. 
Swift/BAT has time in TT, while the standard time system for astronomical sources is Barycentric Dynamical Time (TDB), 
which is TT transformed to the solar system barycenter. 
The maximum difference between the two time systems in the Swift/BAT observations is 0.0016582 seconds, corresponding to a difference in orbital phase of $1.1\times10^{-8}$. 
We ignore the difference in our data analysis. 
 
There is an additional time correction to barycenter from varying light travel time from Her X-1 to Earth throughout the year. 
For the ecliptic latitude of Her X-1 $b = 57.493779 ^{\circ}$, the maximum correction is 
$\frac{\Vec{r} \; \cdot \; \hat{n}}{c} = \frac{(1 \; AU) \; \times \; \cos{b}}{c} = 268.17$ seconds,
with $\Vec{r}$ the vector from barycenter to Earth; $\hat{n}$ is the unit vector from Earth to Her X-1; and $c$ is speed of light.
The result is $\sim 1.8\times10^{-3}$ in units of 1.7-day orbital phase or $\sim 9\times10^{-5}$ in units of 35-day phase. This difference is small enough for
analysis of the 35-day cycle that we do not include it here. 

\subsection{RXTE/ASM Observations}

The Rossi X-Ray Timing Explorer (RXTE) All-Sky Monitor (ASM) \citep{1996ApJ...469L..33L}
monitored the sky in the 2-12 keV energy band. 
The RXTE/ASM lightcurves for Her X-1 were obtained from the Massachusetts Institute of Technology (MIT) RXTE project online database
at the website http://xte.mit.edu/asmlc/ASM.html. 
Time in the RXTE/ASM observations is in units of MJD (TT). 
As for Swift/BAT, the difference in time systems is not important in this analysis and is ignored. 
The same conversion to orbital phases as for the Swift/BAT observations is performed.

The data covers the time period from January 1996 to December 2011 with a total number of 95782 points. 
The sum (2-12 keV) band intensity and uncertainty are used here.
 The bottom panel of Figure~\ref{fig:BATdata} shows a sample of the  RXTE/ASM  lightcurve of Her X-1.
The 35-day cycle is seen in the  RXTE/ASM data, but the data have $\sim$twice 
the noise level as the Swift/BAT data.  

\section{Cross-correlation (CC) Analysis \label{sec:analysis}}

To identify individual 35-day cycles and determine their timing, we use the method of cross-correlation (CC). 
CC measures the similarity of a section of data with a given function or template. 
The CC value is calculated between the template and the data. 
For every data point we use a section of data of length of one 35-day cycle. 
The CC has a local maximum when the data and the function are best aligned in time. 
Thus we use the peaks in CC vs. time to determine the peak times of the 35-day cycle in Her X-1.

Because X-rays from Her X-1 are blocked by its companion star during eclipse, 
the lightcurves contain intervals of sharply lower flux during eclipse.
To avoid these eclipses affecting the cross-correlation analysis, we remove data during eclipse periods prior to the cross-correlation analysis.
We use the orbital ephemeris for Her X-1 from \cite{2009A&A...500..883S}.
Orbital phase zero is defined as mid-eclipse of Her X-1 by its companion Her HZ. 

Eclipse during MH state of Her X-1 lasts from orbital phase 0.935 to 1.065 \citep{1995MNRAS.276..607L}.
Eclipses can last longer because of dips and other extended structures in the system at different 35-day phases,
as shown by {1995ApJ...450..339L} and \citet{2011ApJ...736...74L}.
Thus for the CC analysis we removed all data during eclipse plus an extra 0.095 in orbital phase,
 i.e. we keep data only during the orbital phase interval 0.16 $\le$ phase $\le$ 0.84.
 After the above selection, $67.9\%$ of the original data points remain. 

To avoid having data gaps caused by eclipse removal 
the in-eclipse data were replaced by interpolation between the corresponding edge points in the original observations. 
The values were drawn from a Gaussian distribution,
with mean the average of the two edge points, and standard deviation the sum of the rate errors of the two edge points. 
Rate errors are taken as the standard deviation. 

The previous analysis of RXTE/ASM data by \cite{2010ApJ...713..318L} employed the method of cross-correlation.
This produced a template from RXTE/ASM observations. 
We use their template as input template for first iteration in the CC method, for both the Swift/BAT observations 
and the RXTE/ASM observations. 
Then we use the observations to improve the shape of the template iteratively, as described below.

For this study we define peak of the 35-day cycle to be 35-day phase 0.
We can convert to the standard definition of start of 35-day cycle as X-ray turn-on by subtracting 
the phase difference from turn-on to peak of MH. 
The turn-on has different definitions, but a commonly used
one is the time that the lightcurve reaches 20\% of the peak intensity during the following MH. 
The conversion from 35-day cycle peak time to 35-day cycle turn-on time depends on the shape of the light-curve.
For our final templates for Swift/BAT and RXTE/ASM, turn-on time occurs approximately 0.13 in 35-day phase prior to the peak time.
To avoid having the MH or short high peaks near a boundary in the template function, 
we define the template phase to range from $-0.3 \le \phi_{35d} < 0.7$ with $ \phi_{35d}$ the 35-day phase.

\subsection{Weighted CC Function}

The CC method was introduced for Her X-1's 35-day cycle by  \cite{2010ApJ...713..318L}. 
We modify the method to be less sensitive to data with large relative errors by introducing a weight to the CC function.
The resulting weighted cross-correlation function that we use is: 
\begin{equation}
CC_n \equiv \frac{\sum_{k=0}^{N_n-1} R_k \times T_k \times \omega_k}{\sum_{k=0}^{N_n-1} T_k \times \omega_k} \label{eq:1}
\end{equation}
where $R_k$ is the count rate, $T_k$ is the template value, $\epsilon_k$ is the count rate error and $\omega_k = \frac{1}{\epsilon_k}$
is the weight.

For each data point at time $t_n$, we calculate the $CC_n$ value with the following procedure:
\begin{enumerate}
    \item Find the data point closest to $-0.3 \times t_{35d}$ days previous in time, $t_1$, 
    and the data point closest to $+0.7 \times t_{35d}$ days later in time , $t_2$, 
    with $t_{35d,n}=t_2-t_1$ the length of the 35-day cycle for that piece of data.
    The corresponding number of points is $N_n$ between these two data points. 
  \item For all points $t_k$ in this interval $(0 \le k < N_n)$, convert the times of the data $t_k$ to 35-day phases $\{\phi_k\}$
  using $t_{35d,n}$. 
    \item  With the list $\{\phi_k\}$, we find the corresponding template values $T_k$ by interpolation from the template. 
    We use cubic spline interpolation because the template is relatively smooth. 
    We verified that cubic interpolation works better than linear interpolation.   
    \item We calculate $CC_n$ with equation \eqref{eq:1} for this point. 
\end{enumerate}

After calculating CC values for all data points $n$, we find the peaks in CC values to obtain the times of the 
peaks of the 35-day cycles, $t_{n,peak}$.
These peak times correspond to 35-day phase 0 for each cycle.

To avoid peaks from times when the data is noisy, we only keep peaks in CC value above certain threshold. 
The nominal value for the threshold is $\sim 0.4$ of the globally highest CC value, but it is adjusted manually when there is a clear CC peak above the noise. 
The false peaks due to noise in CC are removed manually using inspection of the CC function and the data. 
From the set of all detected 35-day cycles, we obtain a list of peak times. 
Whenever there are two adjacent peak times we calculate the cycle length as the difference in peak time
 between the current cycle and the subsequent cycle.

To obtain a refined template, we create an average of the detected 35-day cycles for the new template.
To obtain the same time sampling for averaging, we linearly interpolate the data from individual cycles.
Thus the output template from CC is constructed with better phasing than the input template.
Thus, with the output template, the CC function is recalculated to obtain a new set of 35-day peak times, 
new cycle lengths and new template.

This iterative process is carried out until convergence of  peak times, cycle lengths and template.
We compared the difference in peak times and cycle lengths between one iteration and the next iteration.
Convergence is verified by the differences converging to 0. For the Swift/BAT data we used 7 iterations. 

\subsubsection{CC Function for Swift/Bat data}

\begin{figure}[ht!]
 \includegraphics[width=1.0\linewidth]{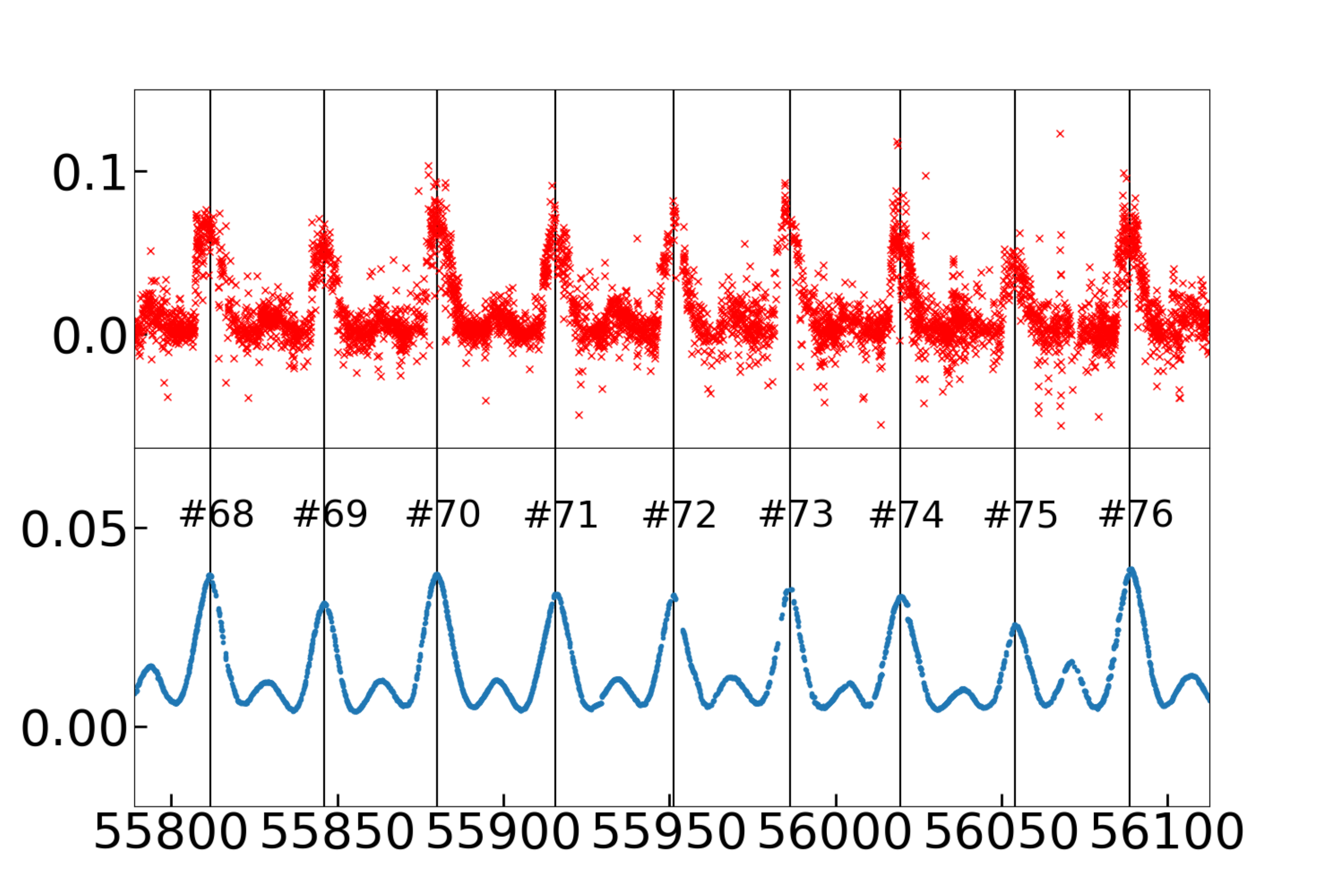}
  \includegraphics[width=1.0\linewidth]{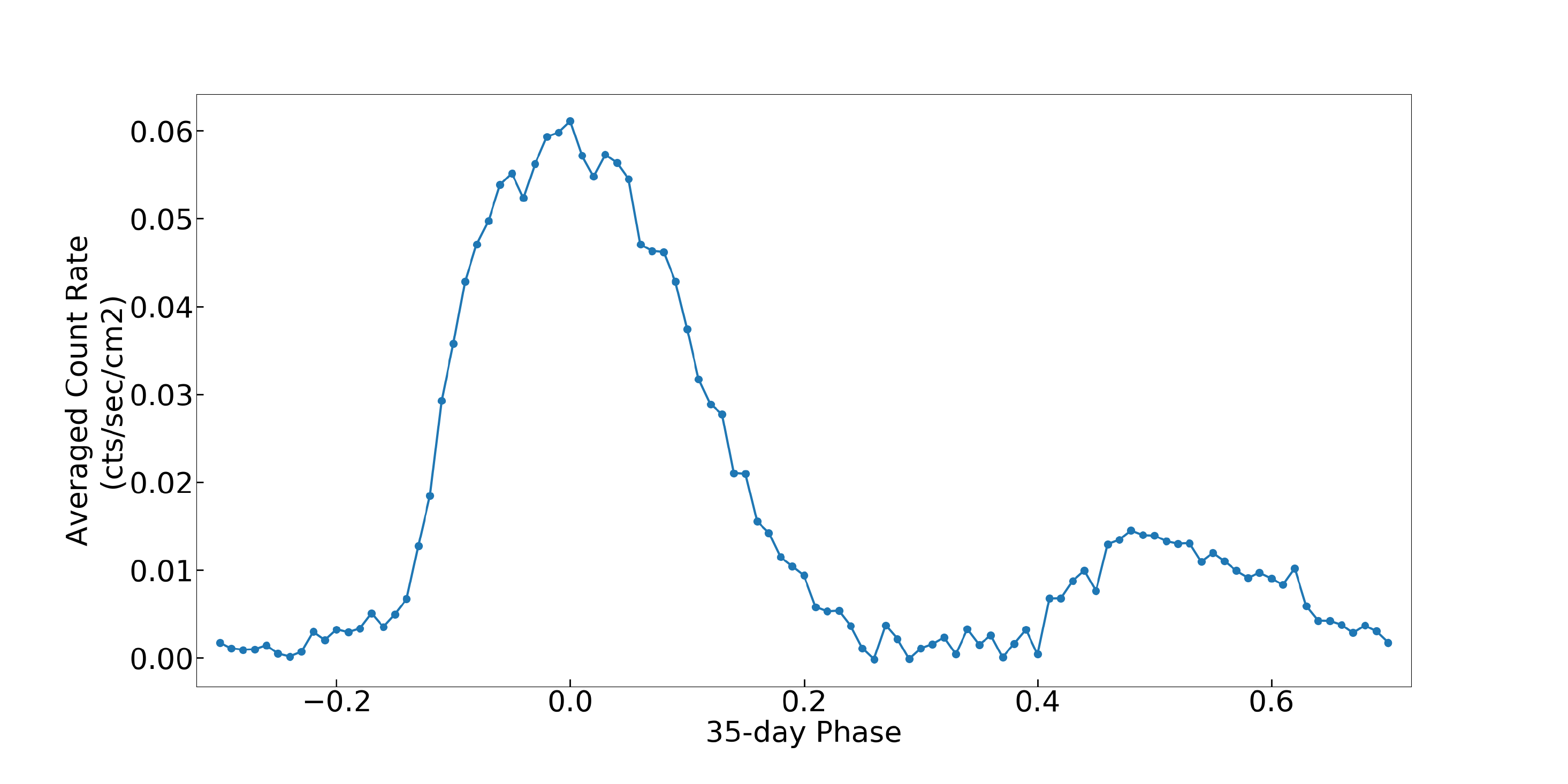}
    \caption{Example Swift/BAT CC for the final iteration of the CC method. 
    The top panel is a sample of the BAT data for the cycles as numbered, in units of ct/cm$^2$/s as a 
    function of time in MJD. 
    The middle panel is CC value vs. time in MJD for those same cycles. 
    The black vertical lines are the CC peaks, one for each 35-day cycle. 
    The bottom panel is the final template, averaged over the 142 cycles with good data coverage. 
    Cycles with significant data gaps are not included in the average.
    \label{fig:BATcc}}
\end{figure}

The top two panels of Figure~\ref{fig:BATcc} show a subset of the data and the corresponding CC function 
from the final CC iteration on the BAT data. 
The final BAT template from the CC analysis is shown in the bottom panel.
As a final step, we linearly interpolate between cycles with peak times measured by the CC analysis 
to estimate peak times and cycle lengths for those cycles which are not clearly detected by the CC analysis.

\subsubsection{CC Function for RXTE/ASM data}

\begin{figure}[ht!]
    \includegraphics[width=1.0\linewidth]{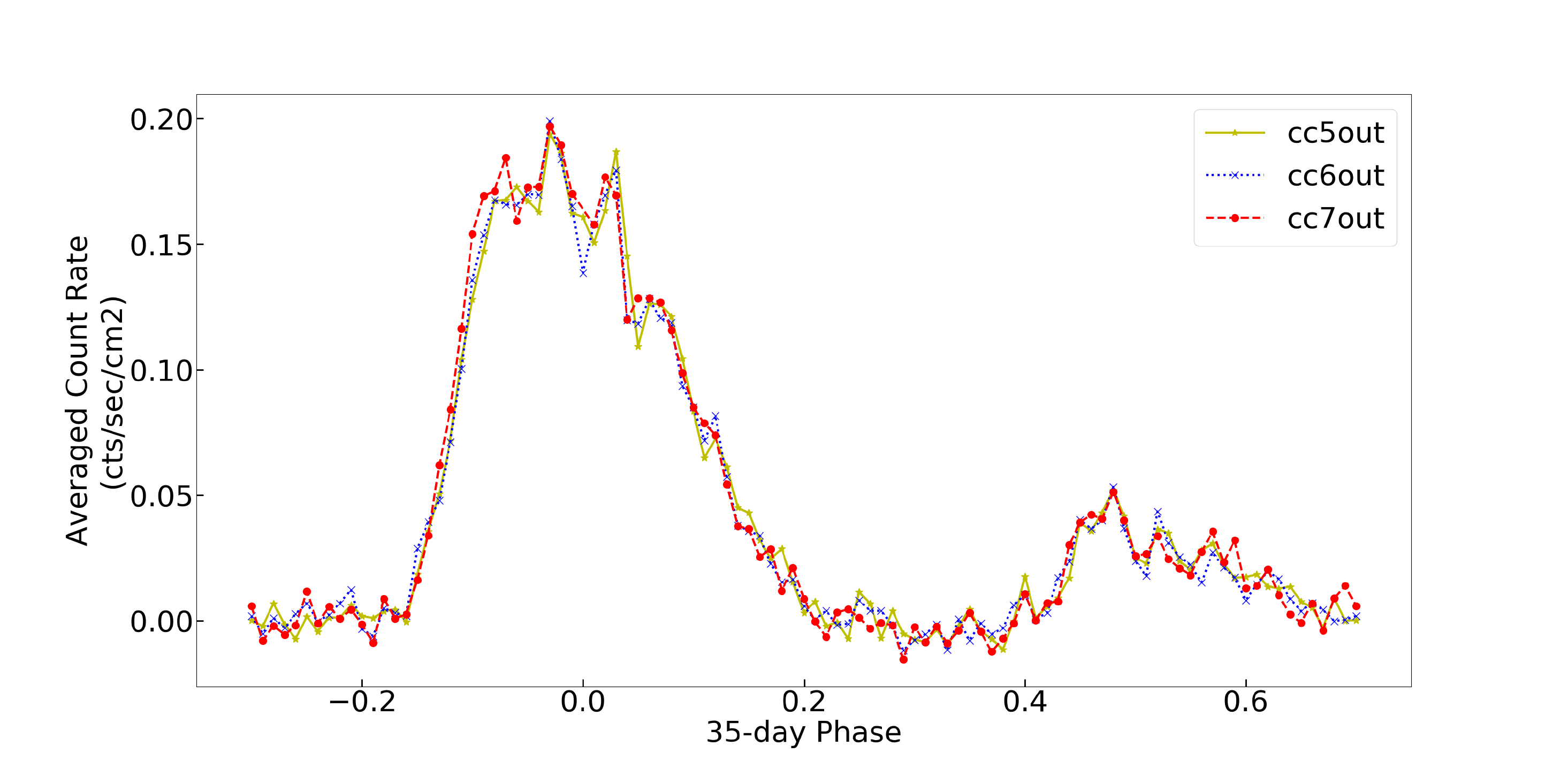}
        \includegraphics[width=1.0\linewidth]{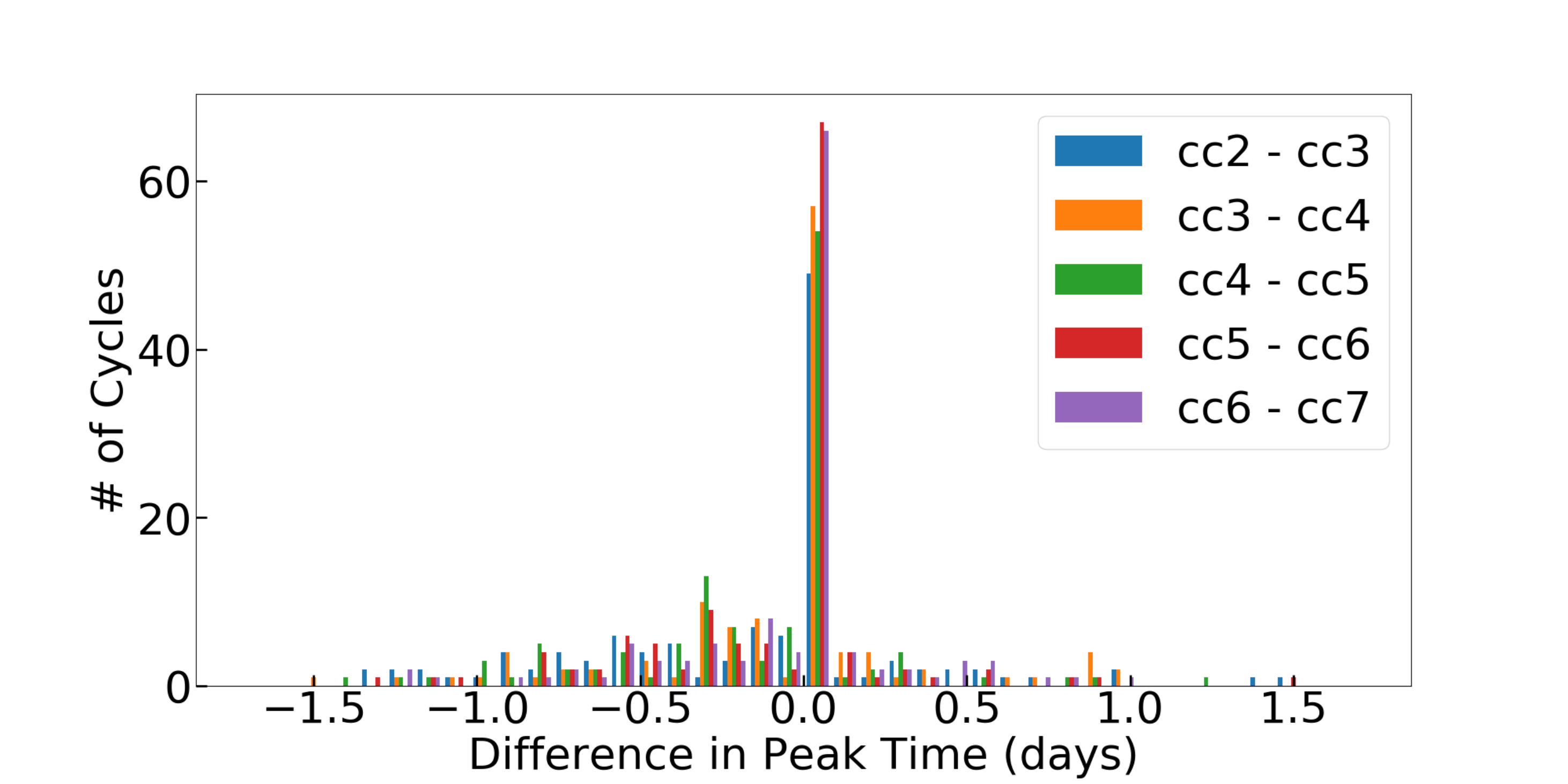}
    \caption{Top panel: 35-day cycle templates for the last three iterations on the ASM data.
    Bottom panel: the change in peak times of 35-day cycles between successive iterations, illustrating the slow convergence.  \label{fig:ASMtemp234}}
\end{figure}

We apply the same analysis to the ASM data. 
In contrast to the results for the BAT data, the results for the ASM data converge for $\sim$4-5 iterations, 
then cease to converge further.
We carried out 7 iterations of CC analysis for the ASM data.
The top panel of Figure~\ref{fig:ASMtemp234} shows the templates for the last three iterations (nos. 5, 6 and 7) from the CC method
applied to the ASM data\footnote{The templates were converted to ct s$^{-1}$ cm$^{-2}$ using an effective area of 30 cm$^{-2}$.}. 
The template changes slightly from iteration to iteration.
The bottom panel shows the histogram of time differences in peak times between successive iterations of the CC method on the ASM data. 
Most cycles agree within a few hours, as shown by the peak with time difference close to 0. 
However, there are a significant number of cycles that do not converge in successive peak times, 
with time differences between -1.5 and +1.5 days. 

We calculated the standard deviation, $SD_{dt}$, of difference in peak times between
iterations. 
 $SD_{dt}$ decreases slowly as more iterations are performed, e.g. between iterations 2 and 7 it decreases from 0.5 days to 0.35-days.
 However the SD of change orbital phase of peak time,  $SD_{d\phi_{orb}}$  does not decrease as iteration number increases. 
 This latter result can be explained from the distribution of change in peak times shown in the bottom panel of  Figure~\ref{fig:ASMtemp234}:
 the relatively constant fraction of cycles that are not measured well (with $dt$ up to 1.5 days) dominate $SD_{d\phi_{orb}}$. 

 Despite the less stable behaviour of the iterations for the ASM data, the differences compared to the previous iteration are within 
 1.5 days for all peak times, and within $\sim0.8$ days for 70\% of the cycles. 
 We tested separating out the cycles with $<0.8$ day differences to perform CC iterations.
 However, we found that the specific cycles with small timing errors changed with each iteration, so
 overall there was not a significant improvement in results. 
  The results in Section~\ref{sec:discussion} are based on all CC-detected cycles from ASM data. 
 
\section{Results and Discussion \label{sec:discussion}}
 
The outputs of the CC analyses on the BAT and ASM data sets are a list of 35-day cycle peak times 
and a list of 35-day cycle lengths for each of BAT and ASM. 
These are presented as online tables, with the first several entries given here in Table~\ref{tab:BATcycles} for the BAT data and in 
 Table~\ref{tab:ASMcycles} for the ASM data.
 
A total of 146 CC peaks are detected in BAT observations. 
We add the cycles with weak CC peaks manually and label them ``1" to be distinguished from algorithm detected ones (labelled 0). 
A few cycles have data missing near 35-day peak; we label these ``2" and provide a time range. 
These cycles have their cycle numbers duplicated, and the cycles right before them are also duplicated to show the range of cycle lengths. 
For the remaining cycles with missing data in most of the cycle, we divide the lengths in time with number of cycles in the time period and 
calculate approximate MH peak times and cycle lengths. 
In these cases the orbital phases are not applicable, so we assign ``-1" in the orbital phase column of Table \ref{tab:BATcycles}). 
They are also labelled ``1". 

\subsection{Cycle Lengths}

\begin{figure}[ht!]
    \includegraphics[width=1.0\linewidth]{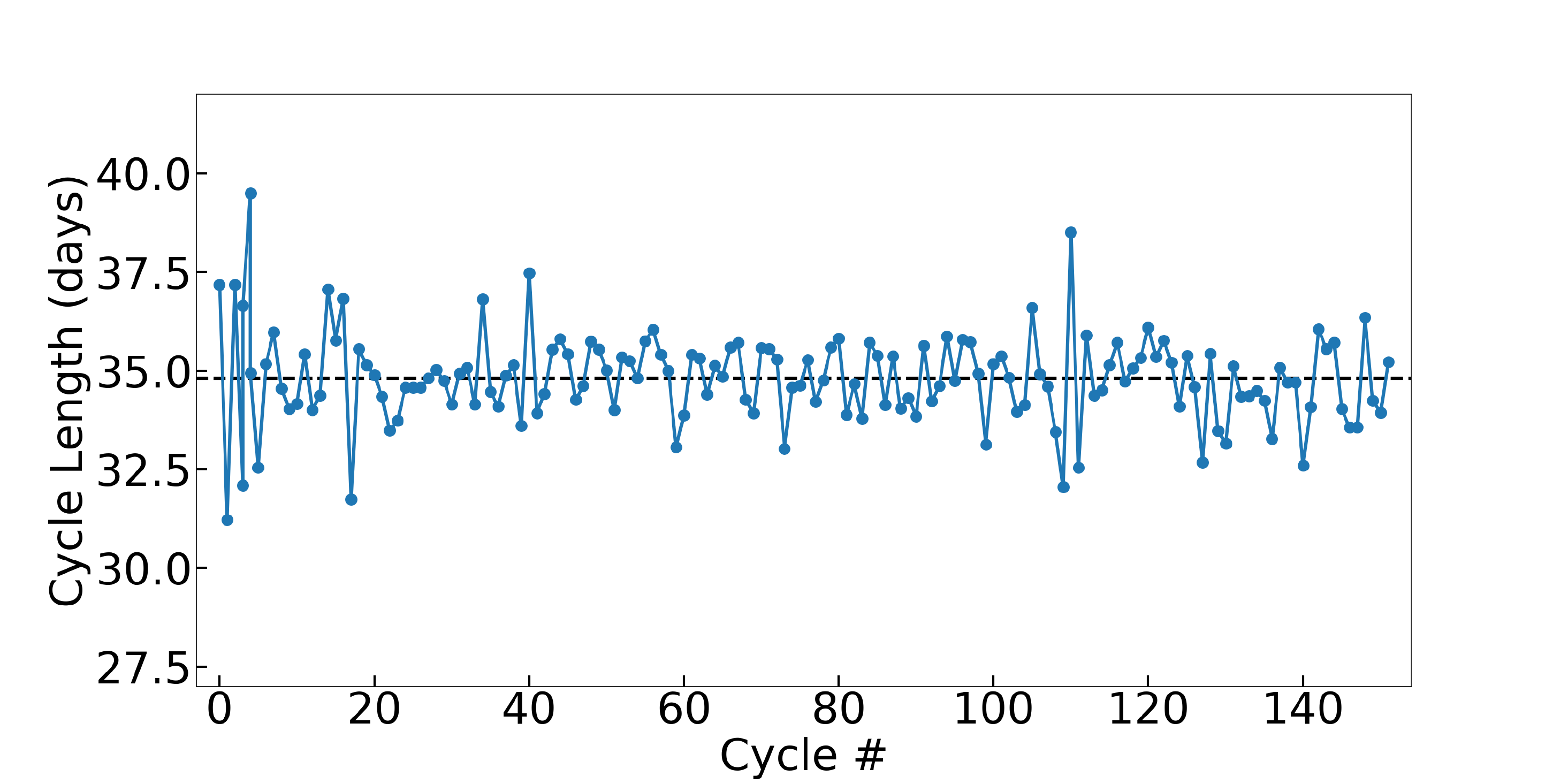}
        \includegraphics[width=1.0\linewidth]{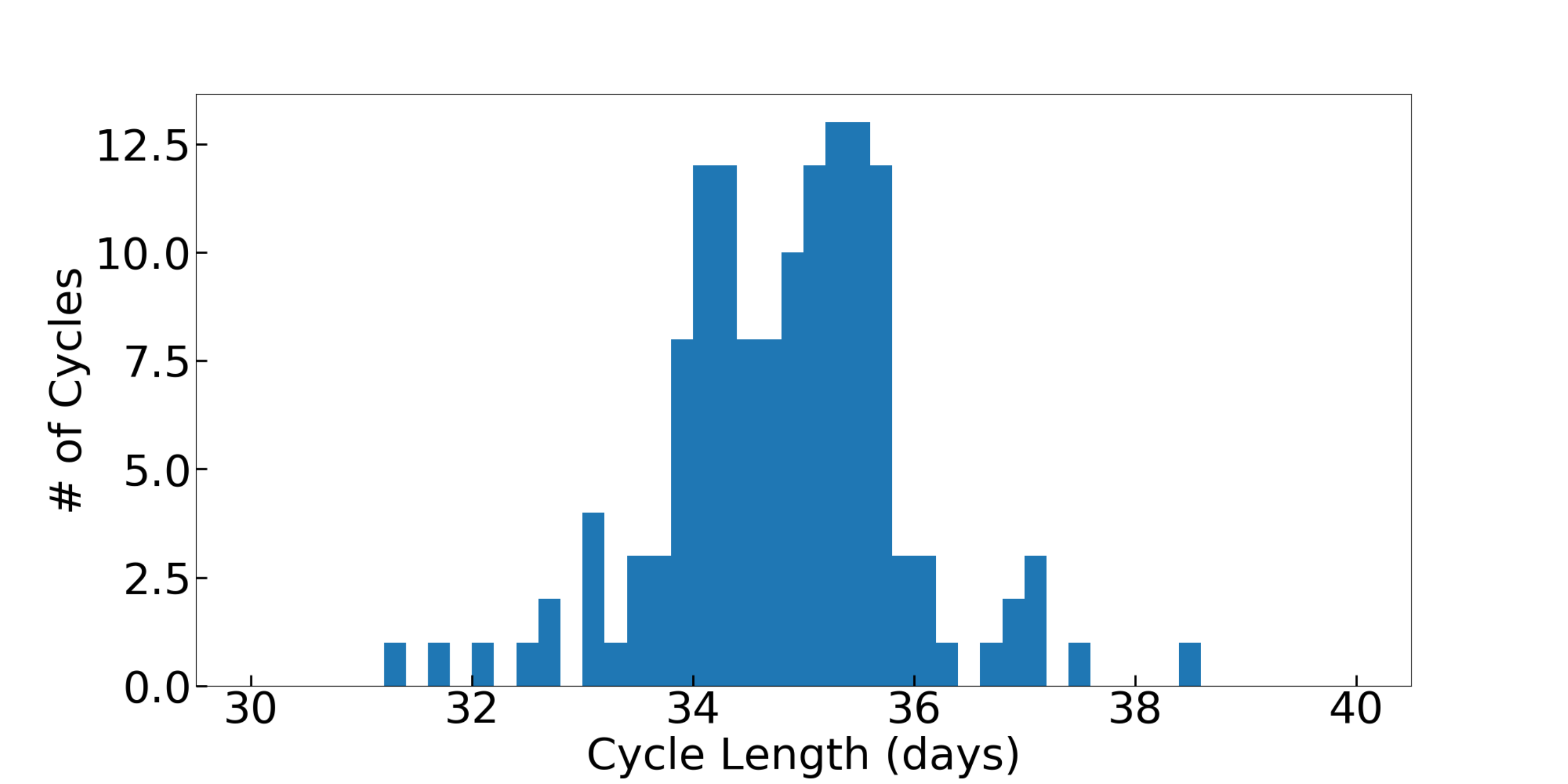}
    \caption{Swift/BAT cycle lengths: The top panel shows cycle length vs. cycle number and the bottom panel shows the distribution of cycle lengths.
    The first cycle has its peak in March 2005 (MJD 53438) and last cycle peak in July 2019 (MJD 58692). 
    The black dashed line in the top panel shows the mean 35-day cycle length (34.787 days).    \label{fig:BATcyc}}
\end{figure}

The top panel of Figure~\ref{fig:BATcyc} shows the cycle lengths vs. cycle number for the BAT data.
There is no clear change of the 35-day cycle length with time. 
The lower panel shows the histogram of cycle lengths measured using the BAT data. 
The top panel includes all cycles (CC-detected and weak cycles), whereas the bottom panel only contains CC-detected cycles. 


\begin{figure}[ht!]
    \includegraphics[width=1.0\linewidth]{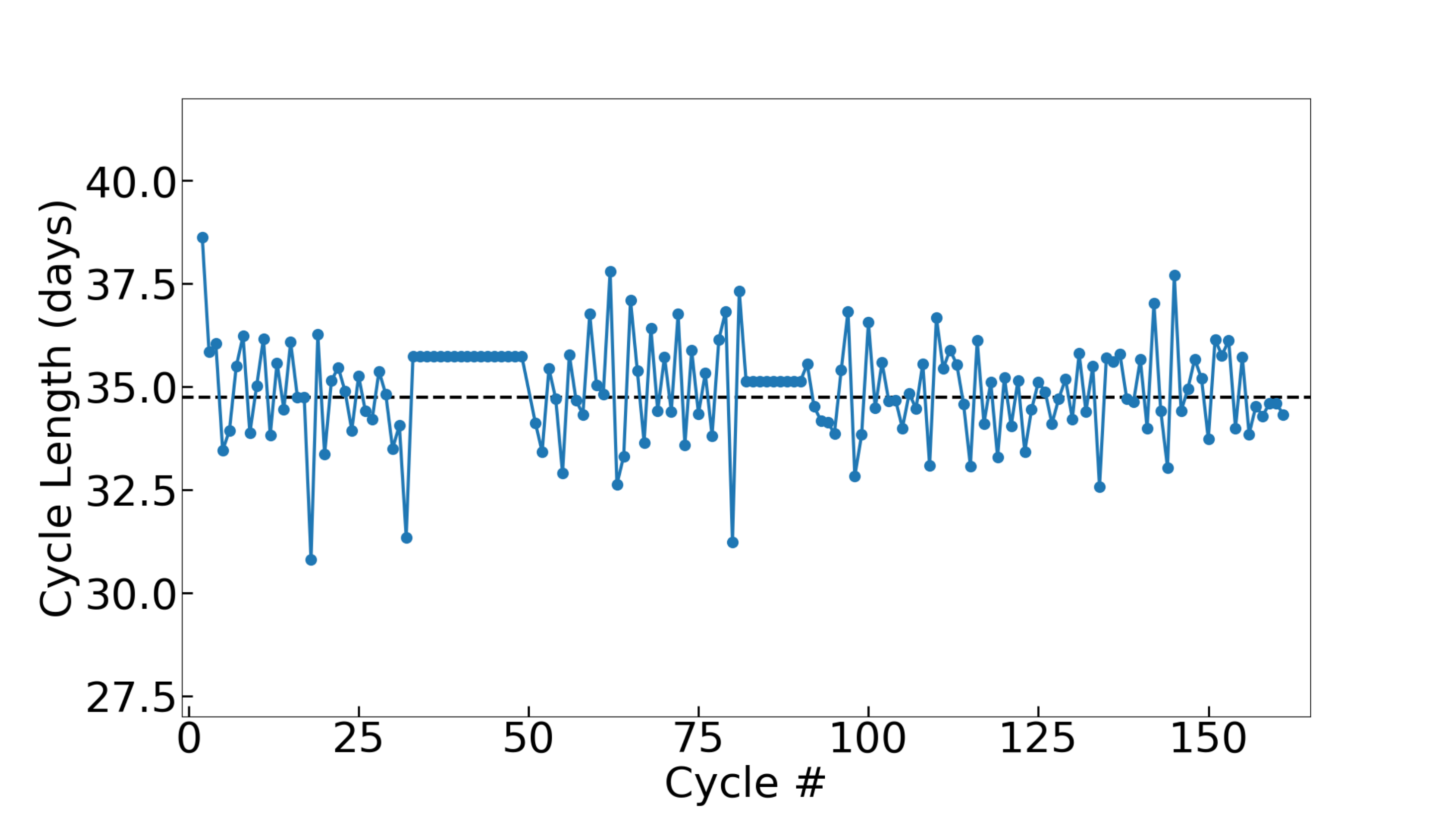}
        \includegraphics[width=1.0\linewidth]{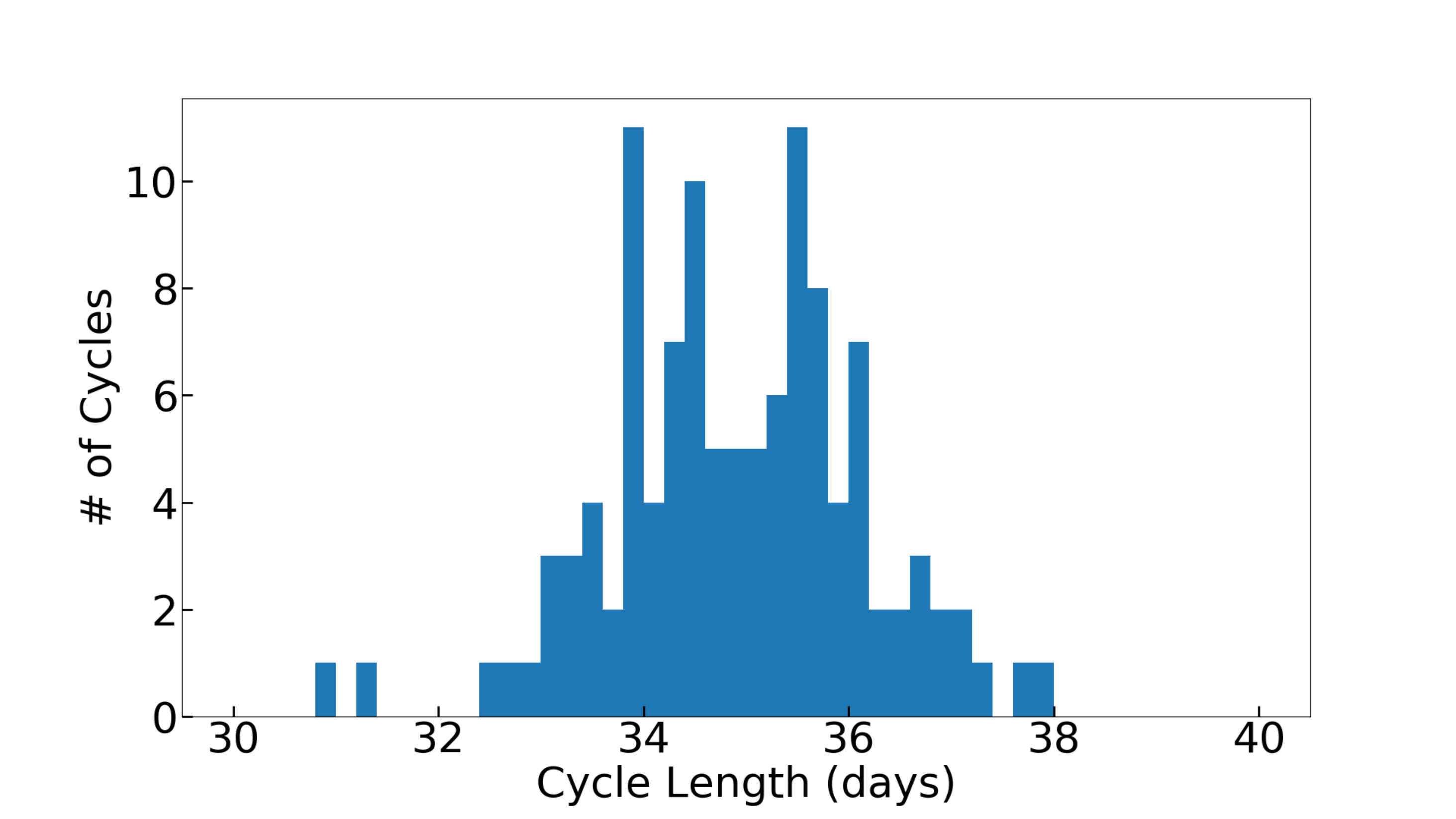}
    \caption{RXTE/ASM cycle lengths: The top panel shows cycle length vs. cycle number.
    The bottom panels shows the distribution of cycle lengths for CC-detected cycles. 
    The first cycle has its  peak in March 1996 (MJD 50146) and last cycle peak in May 2011 (MJD 55705). 
    The black dashed line in the top panel shows the mean 35-day cycle length  (34.553 days).   
    \label{fig:ASMcyc}}
\end{figure}

The top panel of Figure~\ref{fig:ASMcyc} shows the cycle lengths vs. cycle number for the ASM data.
As for the BAT analysis, there is no clear change of the 35-day cycle length with time. 
 The anomalous low states are visible as the two periods of constant interpolated cycle length, where no CC-detected cycles are found.
The lower panel shows the histogram of cycle lengths measured using the ASM data. 
The top panel includes all cycles (CC-detected and weak cycles), whereas the the bottom panel only contains CC-detected cycles. 
We compared the distributions of cycle lengths from the last 3 CC iterations. 
Iterations 5, 6 and 7 give the same distribution of cycle lengths within Poisson errors per bin. 

Table \ref{tab:BAT-ASMcyclen} gives the average cycle length and standard deviation, SD, from the BAT CC analysis
 for the ASM CC analysis, including the CC-detected cycles (``CC"). 
 The numbers are nearly the same ($<0.1$d different for BAT and $<0.2$d different for ASM) if we include the interpolated cycle lengths, which were not measured by the CC analysis.
    
\subsection{35-day cycle peak and relation to Turn-on}

Here we investigate the question of the timing of turn-on relative to 35-day cycle peak
using the results from our CC-analysis of the BAT data and of the ASM data.
Turn-on is defined as the time prior to MH peak when the flux reaches 20\% of the peak flux. 
We measured the 35-day cycle phase difference between turn-on and peak using the final template lightcurves 
for BAT and ASM, and obtained 0.13 for BAT and 0.11 for ASM.

\begin{figure}[ht!]
    \includegraphics[width=1.0\linewidth]{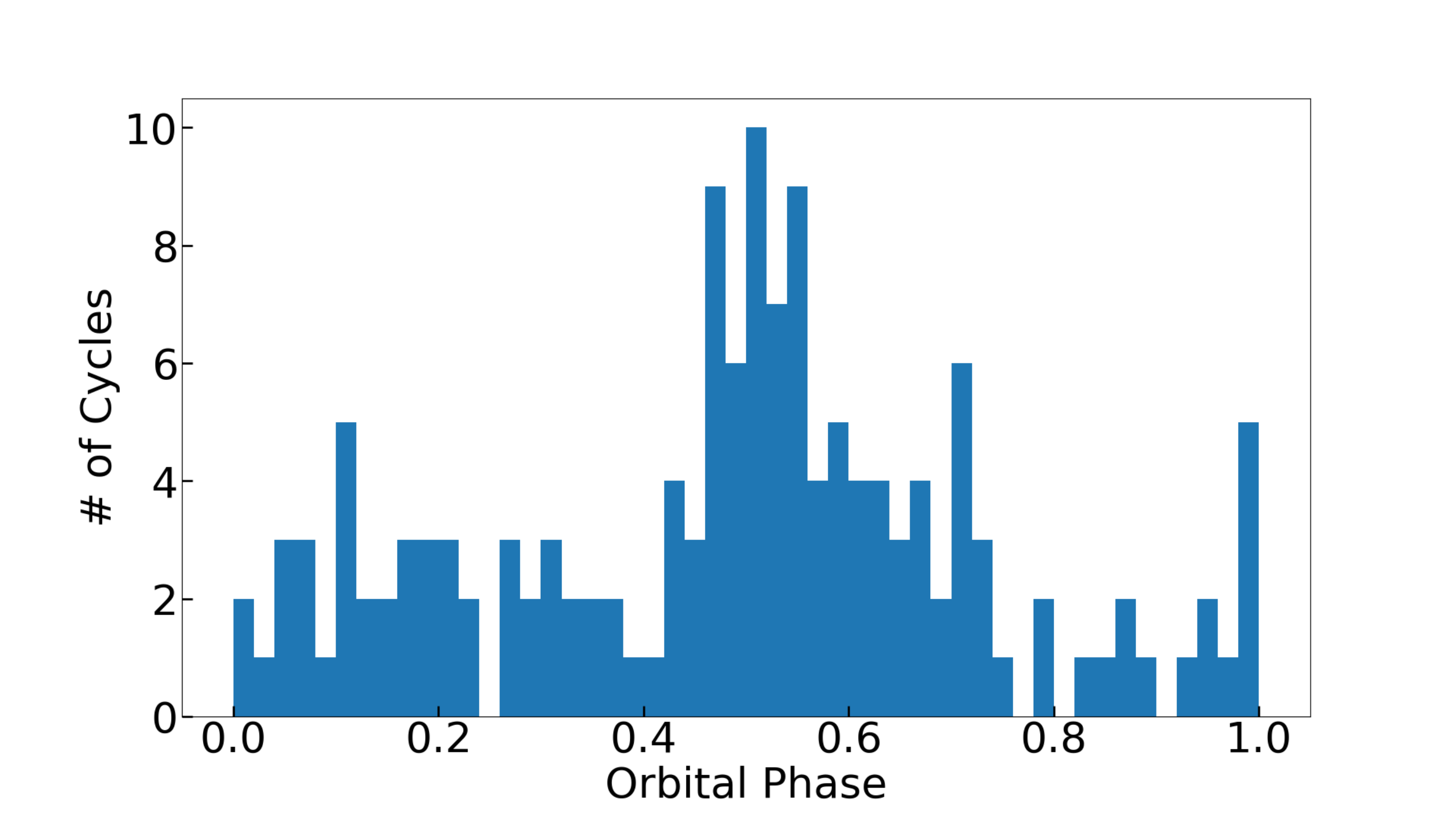}
     \includegraphics[width=1.0\linewidth]{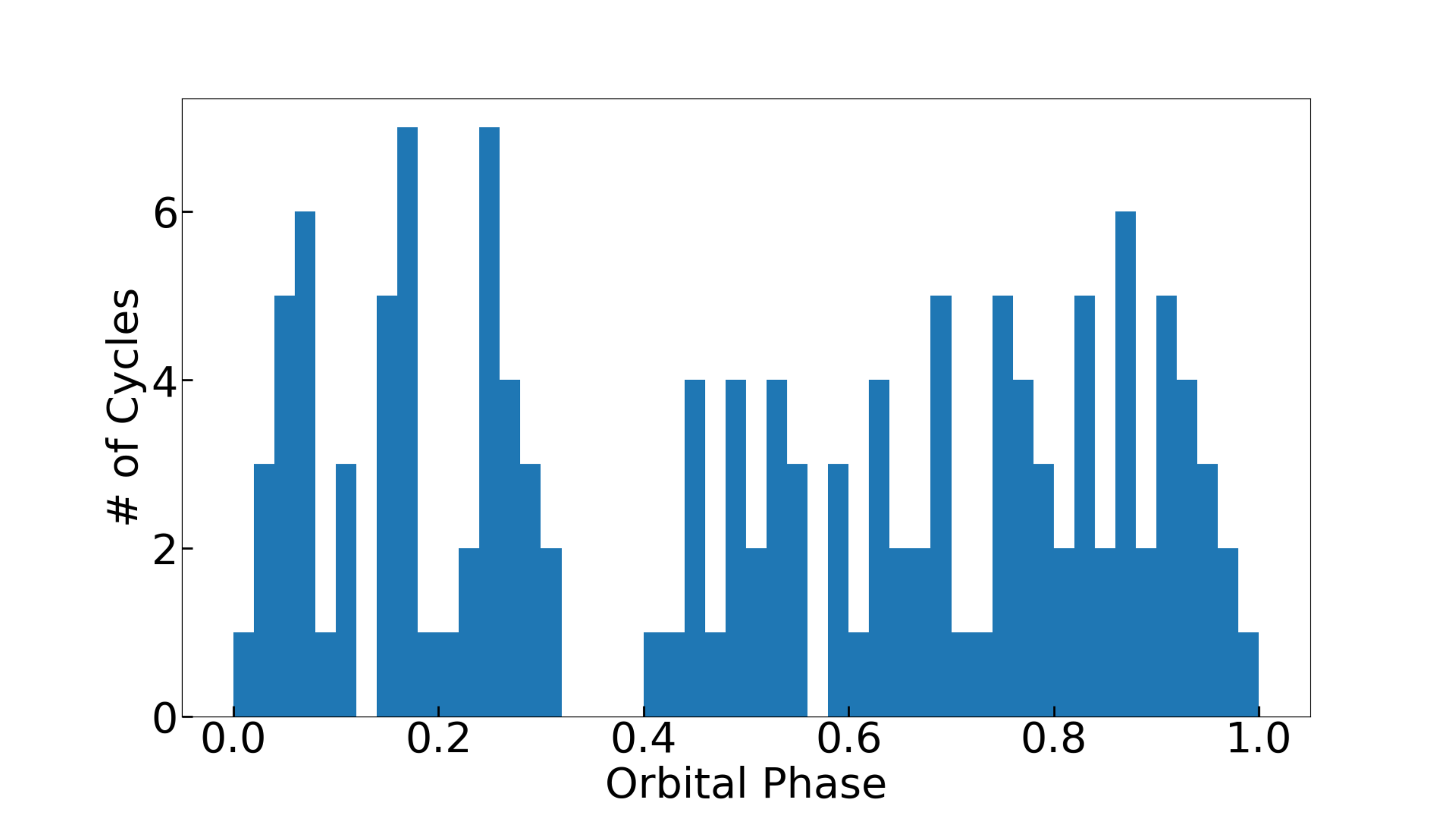}
    \caption{Top panel: distribution of orbital phases of 35-day cycle peaks from CC analysis of BAT data.    
    Bottom panel: distribution of orbital phases of 35-day cycle peaks from CC analysis of ASM data.
    \label{fig:BATorb_hist}}
\end{figure}

The distribution of orbital phases of 35-day cycle peaks is shown in Figure~\ref{fig:BATorb_hist} from 
the CC analysis of BAT data (top panel) and from ASM data (bottom panel). 
For both cases only CC-detected cycles are shown.
For BAT cycles, there is a clear peak near orbital phase 0.5. 
The distribution is consistent with the sum of a uniform distribution  and a Gaussian distribution.
The uniform distribution accounts for 76\% of the total probability. The Gaussian accounts
for the remaining 24\% and has a mean phase of 0.512 with $\sigma$=0.048.
For ASM cycles, the distribution of orbital phases of 35-day cycle peaks is consistent with a uniform distribution.

The orbital phase of 35-day turn-on has been a subject of interest in the literature. 
Early references often found turn-on to preferentially occur near orbital phase 0.2 or 0.7 (e.g. 
\citealt{1999ApJ...510..974S} and references therein).
In contrast, \citet{2010ApJ...713..318L} analyzed orbital phases of turn-on observed 
with RXTE/ASM and found a uniform distribution.
Here we convert the BAT-determined 
0.13 35-day phase difference, using the average cycle length, to 4.52 days or 2.66 
binary orbits. This converts the 0.51 orbital phase for peak to 0.85 orbital phase for turn-on.   
In summary, we are finding $\sim$76\% of the cycles have no preferred orbital phase for turn-on,
and $\sim$24\% of the cycles are clustered around orbital phase 0.85.

\subsection{Comparison of BAT and ASM 35-day Templates}

The BAT data for Her X-1 covers the energy band 15-50 keV, whereas the ASM data covers the energy band 2-12 keV.
The Her X-1 spectrum during different parts of the 35-day cycle is variable and well studied. 
A recent summary of the spectrum of Her X-1 is given in \citet{2019ApJ...871..152L} (and references therein).
Summaries of spectral variations  in terms of changing softness ratios are given by \citet{2011ApJ...736...74L} (and references therein).
For example, Fig. 4 of \citet{2011ApJ...736...74L} shows the 2-4 keV to 9-20 keV softness ratio vs. 35-day phase.
The spectrum is softest at $\simeq0.3-0.4$ during MH state and harder at $\simeq0.2-0.3$ during short high state.
The softness ratio is $\simeq0.25$ during the low states, similar to that for short high state. 
Thus we expect differences in 35-day cycle template shape between BAT 
and ASM.

\begin{figure}[ht!]
    \includegraphics[width=1.0\linewidth]{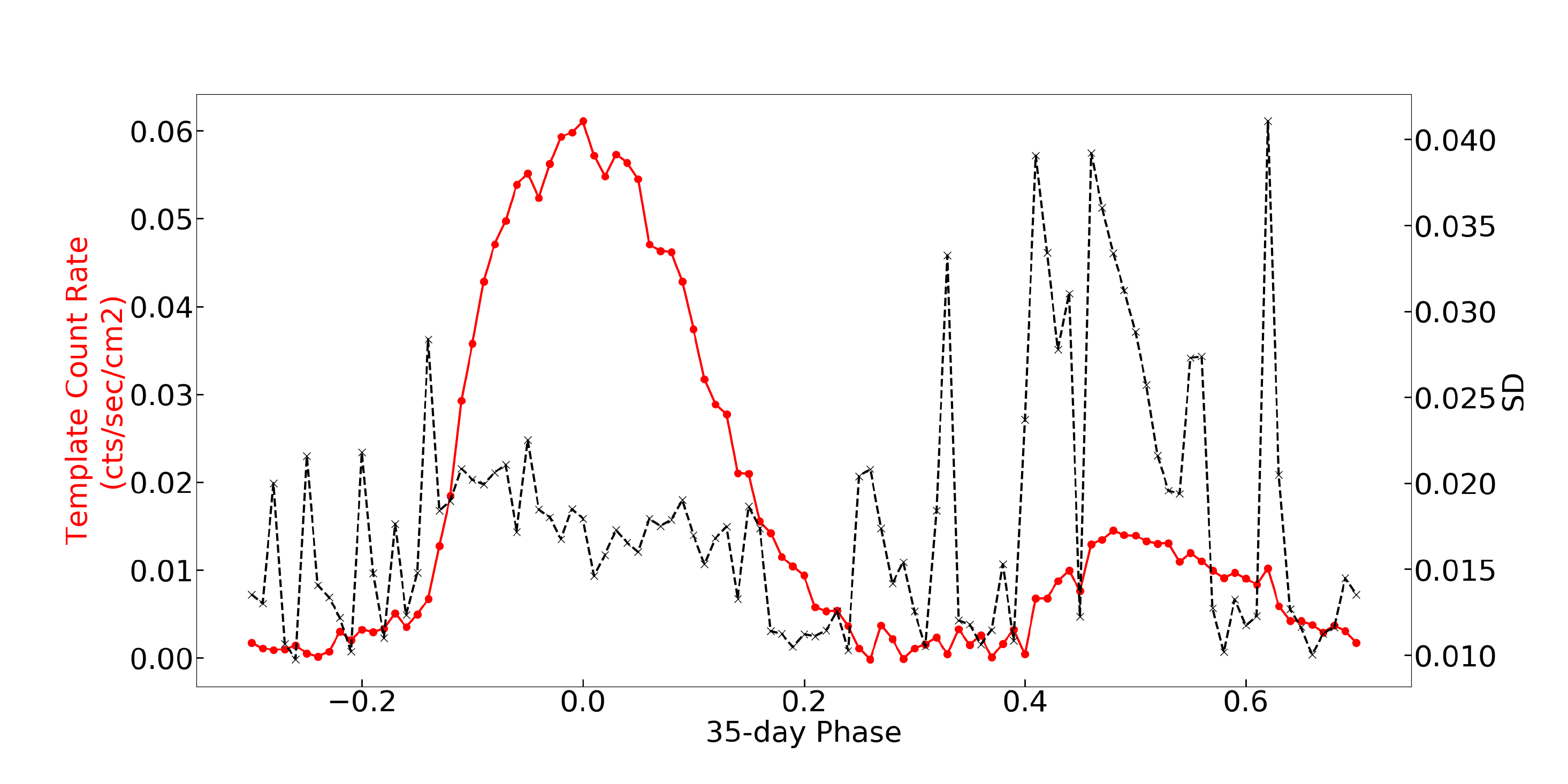}
        \includegraphics[width=1.0\linewidth]{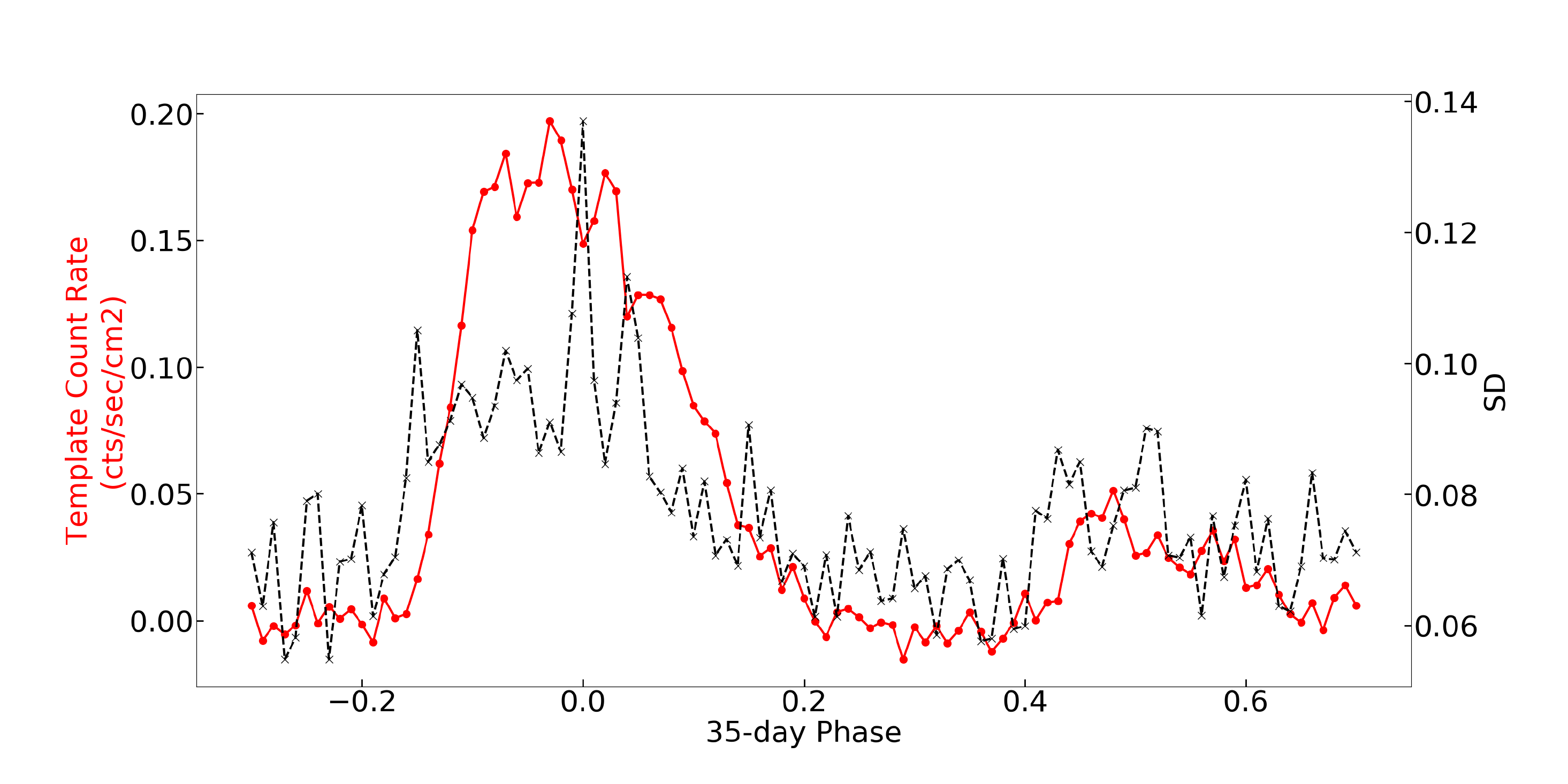}
    \caption{Top panel: BAT 35-day cycle template (red circles) and standard deviation per bin (SD, black crosses) with average SD $0.018$ ct cm$^{-2}$s$^{-1}$ for 142 CC-detected cycles.  
    Bottom panel: binned ASM template (red circles) and standard deviation per bin (SD, black crosses)
 with average SD $0.077$ ct cm$^{-2}$s$^{-1}$ for 115 CC-detected cycles.  
     \label{fig:BATtemp}}
\end{figure}

Figure~\ref{fig:BATtemp} compares the shapes of the templates output by the CC analysis for BAT (top panel) and for ASM (bottom panel). 
The SD in rates from different cycles is shown by the crosses and dashed line for both BAT and ASM templates.
Overall the BAT SD has values similar to the BAT count rates, and the ASM SD is similar to the ASM count rates.
The BAT (high energy) template shows lowest SD during low states, intermediate SD during MH state, and highest SD during short high state.
The ASM (low energy) template shows lowest SD during low states, intermediate SD during short high state, and highest SD during MH state.
I.e., the high energy cycle-to-cycle variability is largest for short high state, 
and the low energy cycle-to-cycle variability is largest for MH state.
At both low and high energy, the cycle-to-cycle variability is smallest for low states.

The disk blocks the direct emission from the neutron star during low states 
(e.g. \citealt{2003MNRAS.342..446L} and  \citealt{2002MNRAS.334..847L})
and only allows the observer to detect X-rays scattered from an
extended region. Thus low variability for the low states is not surprising. 
Her X-1 spents a high fraction of its time in dips during short high compared to MH \cite{2011ApJ...736...74L}.
The MH spectrum has the least amount of absorption, so absorption dips make MG more variable at low energy than they do for short high.
This is consistent with the results shown in Figure~\ref{fig:BATtemp}. 
The larger variability at high energy for short high state is likely caused by variability in the scattering of hard X-rays.
The scattering is from the compact the inner edge of the disk, as shown by the models of \citet{2003MNRAS.342..446L}.

\subsection{Swift/BAT and RXTE/ASM comparison during 2005-2011}

\begin{figure}[ht!]
\plottwo{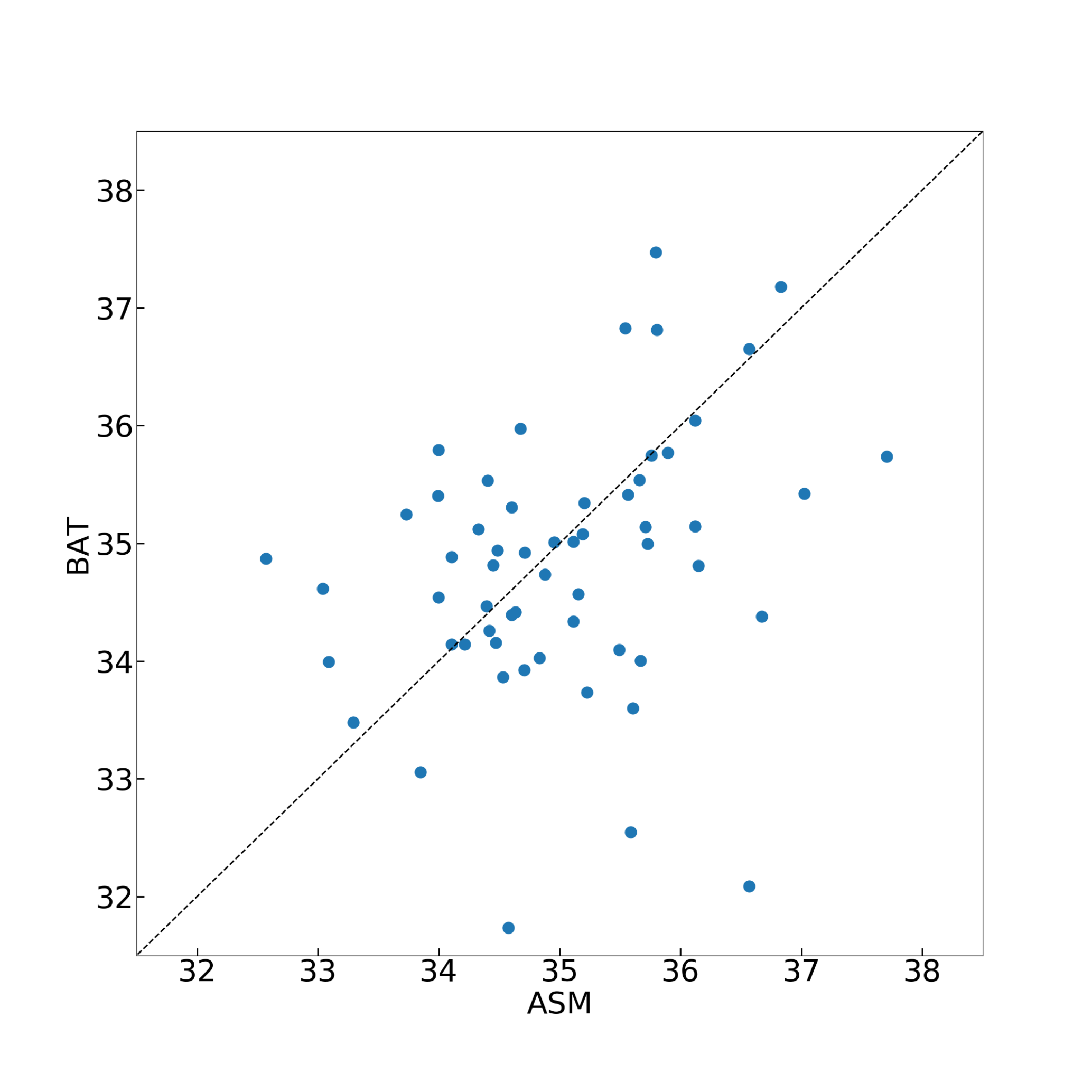}{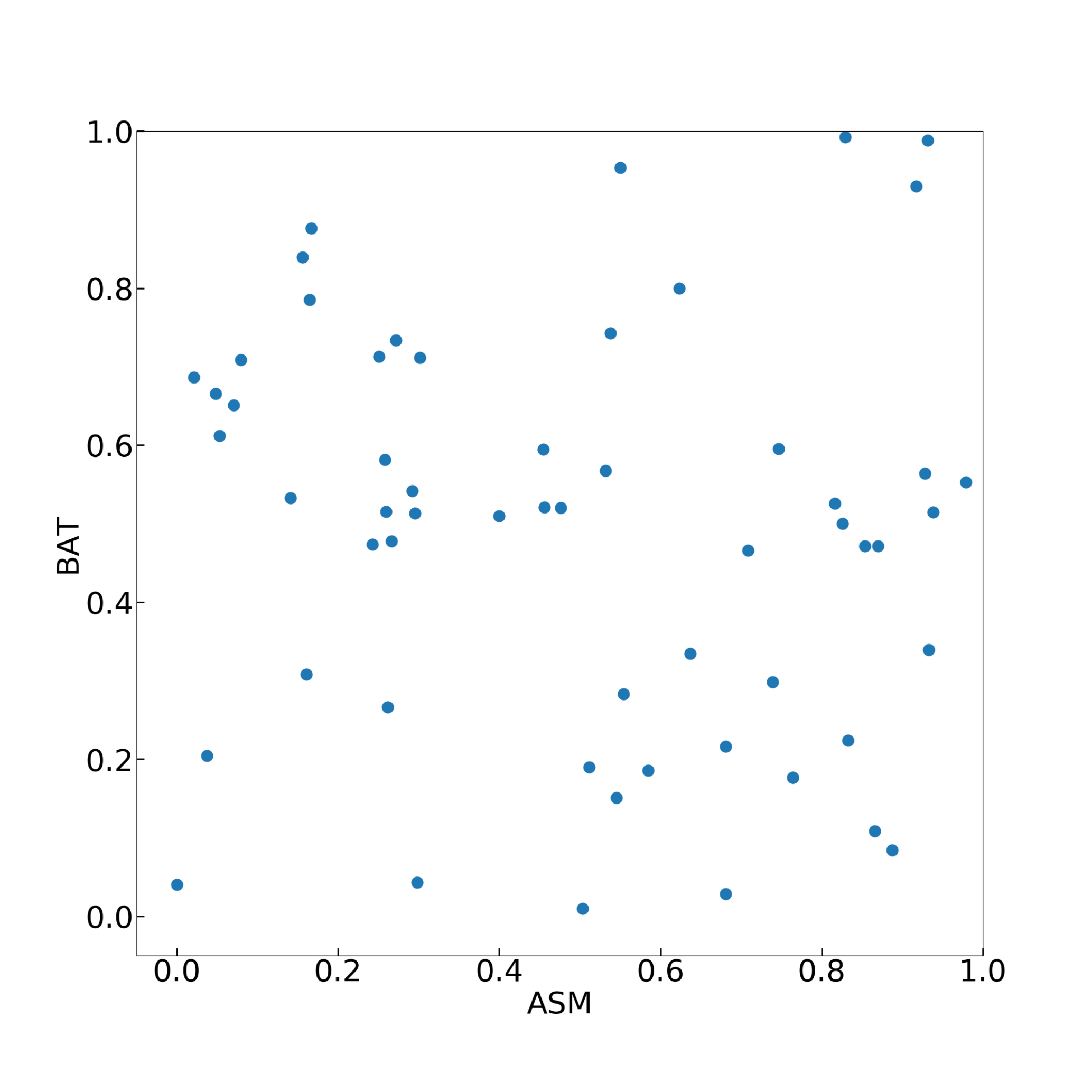}
    \caption{Left panel: comparison of cycle lengths derived from peak times for BAT and for ASM.
    This includes only cycles detected in both BAT and ASM data. Right panel: comparison of orbital phases of peak times for BAT and for ASM.  \label{fig:BAT-ASMovlp_cyc_len}}
\end{figure}

\begin{figure}[ht!]
\plotone{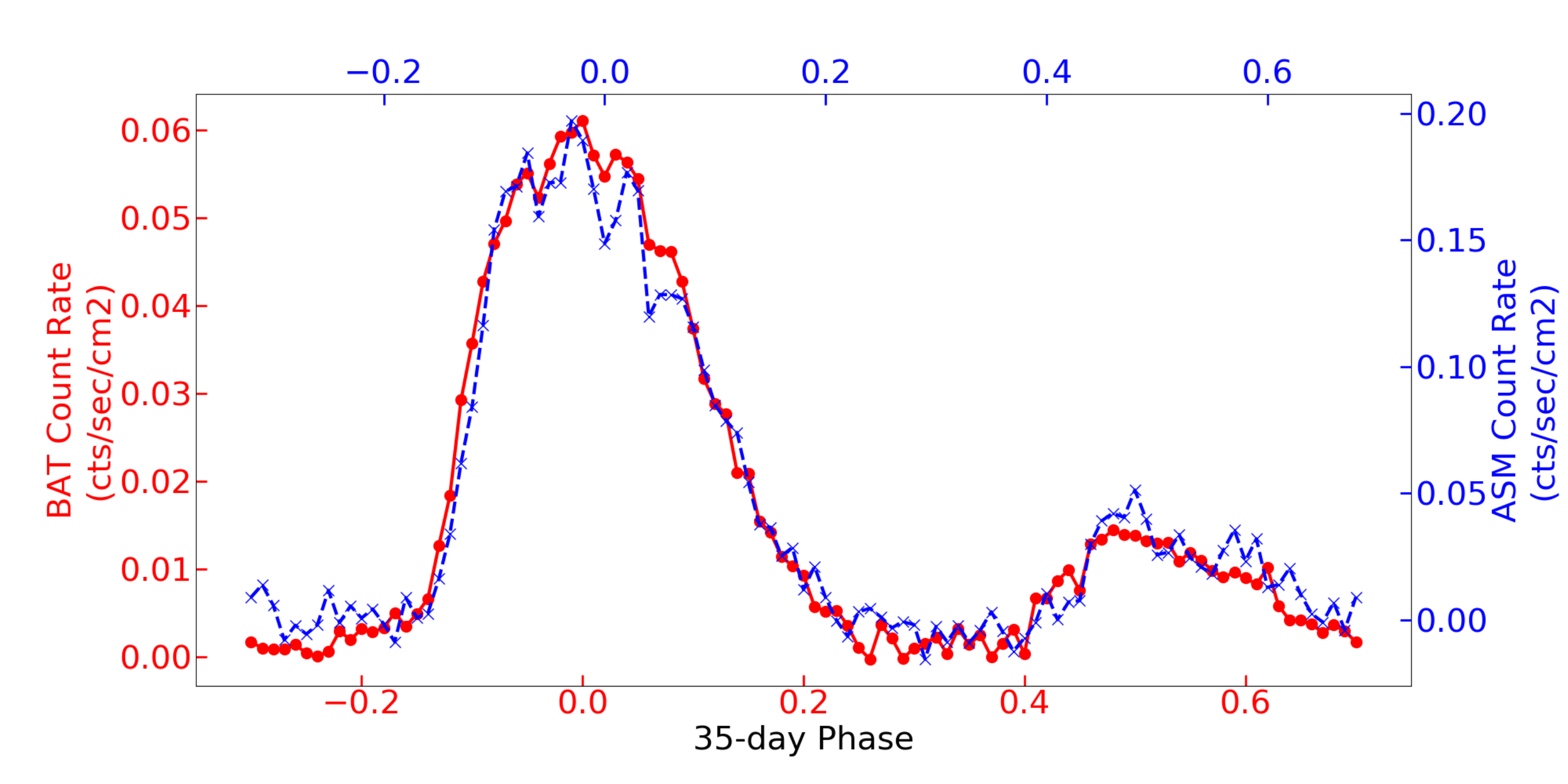}
    \caption{BAT and ASM 35-day cycle templates aligned in 35-day phase 
    using an offset of 0.02 in 35-day phase.
    This shows the BAT rise of MH occurs earlier than for ASM, and the declines from MH 
    occur at the same time.  \label{fig:BATASMalign}}
\end{figure}

There is an overlap in time for the Her X-1 observations for BAT and ASM. 
The overlap is from March 9th, 2005 (MJD 53438), to May 24th, 2011 (MJD 55705), or 2267 days (6.21 years). 
During this overlap period, there are 63 maxima in BAT data and 58 in ASM identified by CC. 
Differences between ASM and BAT in the overlap period are calculated from all cycles (not only CC detected ones), including weak ones and calculated ones. 
Of these, 59 agree within one orbital period (1.7 days) of Her X-1. 
The mean periods and SD for the overlap period are listed in Table~\ref{tab:BAT-ASMcyclen} 
and the time differences (ASM-BAT) for each cycle are listed in Table \ref{tab:BATASMcycles}.
The difference in 35-day cycle peak time between BAT-detected and ASM-detected cycles ranges between $\sim-1.5$ and $\sim+1.5$ days.

Figure~\ref{fig:BAT-ASMovlp_cyc_len} shows 
 the comparison between BAT and ASM cycle lengths (left panel) and between BAT and ASM derived orbital phases of peak (right panel). 
 The error in times of peak for ASM is estimated to be $\sim\pm 1$ day from the bottom panel of  
 Figure~\ref{fig:ASMtemp234}. 
 This error is enough to yield the scatter in peak times seen in the left panel of  
 Figure~\ref{fig:BAT-ASMovlp_cyc_len}, and to fully mix the orbital phases of peak as seen in the right panel.
The conclusion is that the errors in peak time measured with ASM are large enough that the distribution
of orbital phases of peak times with ASM does not give useful information.

From the data during the overlap period, the average time difference (ASM-BAT) in peak of the 35-day cycle is 
$0.366$ days (0.011 in 35-day phase) and its SD is $0.745$ days  (0.022 in 35-day phase).
Thus the 35-phase difference between ASM and BAT derived from the overlap period is $0.01\pm0.02$. 

 Figure~\ref{fig:BATASMalign} shows average BAT and ASM 35-day cycle templates aligned using an offset
 of 0.02. This produces a good alignment of the declines from MH peak over 35-day phase 0.1 to 0.2,
 and of the rises to MH over 35-day phase -0.15 to -0.10, and is consist with the phase difference
 measured from the overlap period.
It is seen that the 3 dips in the templates, one just before MH peak and two after, 
are simultaneous in the BAT energy band and in the ASM energy band. 
This is confirmation that the alignment of the two templates is correct.
The rises and declines for Short High are also seen to occur at the same time for ASM and BAT.

We compared the histograms of orbital phase of peaks of the 35-day cycle for BAT and ASM from the overlap period. 
Each histogram is consistent with the corresponding histogram shown in Figure~\ref{fig:BATorb_hist} for the full data sets.
I.e. the BAT histogram of orbital phases is consistent with a uniform plus Gaussian component centered near orbital phase 0.5,
 the ASM histogram of orbital phases is consistent with a uniform distribution, and the ASM distribution is not consistent with the BAT distribution.
As for the full sets of 35-day cycles, this result for the overlap time period can be explained 
by the larger uncertainty in the ASM measurements of peak times of the 35-day cycles.

The most complete previous analyses of 35-day turn-on times were by \citet{2010ApJ...713..318L} 
 and \citet{2013A&A...550A.110S}.
The first study gave a table of 147 turn-on times, found no systematic change in cycle length with time (their Fig. 5), 
and found a uniform distribution for orbital phases of turn-on (their Fig. 6).
The latter study gave ``Observed - Computed'' turn-on times, which depend strongly on the Computed model, vs. MJD (their Fig.2) 
from analysis of the same data plus some earlier observations.
If the mean 35-day cycle length  of 34.9 d from \citet{2010ApJ...713..318L} is used, then the
cycle lengths of the two studies are consistent for the time period covered by ASM observations.
Here we confirm the uniform distribution for orbital phases of turn-on by re-analyzing the ASM data. 
However we attribute the result to the uncertainty in measuring the time of peak (or turn-on) of 35-day cycles. 

With the BAT data, the uncertainty in measurement of peak time is
smaller and we detect a maximum in the orbital phase distribution at orbital phase of 0.512. 
76\% of the cycles have a uniform distribution, and 24\% of cycles show the peak in orbital phase.
We convert from time of peak of the 35-day cycle and to time of turn-on by subtracting the orbital 
phase difference of 2.66 orbits between turn-on and peak.
The result is a maximum in turn-on time distribution at orbital phase 0.85 with a spread $\sigma=$0.05.

\section{Summary and Conclusion \label{sec:conclusion}} 

Here we analyze long term observations of Her X-1 with the Swift/BAT instrument and the RXTE/ASM instrument.
These instruments have monitored the flux from Her X-1 in the energy bands 15-50 keV (BAT) and 2-12 keV (ASM)
over the long time period of $\simeq$14.5 years for BAT and for ASM.
The measurements have a cadence of a few per day.
With this extensive monitoring, we carry out a study of the 35-day cycle in this system.

The main results are as follows. 
The ASM data includes two periods where there are anomalous low states (ALS) for Her X-1.
ALS were discovered by \cite{1985Natur.313..119P} and a recent analysis is given by \cite{2010ApJ...715..897L}.
The first ALS is from $\simeq$MJD51223 to MJD51831 with 17 or 18 missing 35-day cycles;
the second is from $\simeq$MJD52915 to MJD53232 with 9 cycles missing.
The BAT data includes no periods of ALS: every cycle is seen in the data, although some are weak.

The BAT data has lower noise than the ASM data, as seen in the sample lightcurves shown in Figure~\ref{fig:BATdata}.
The CC method yields a good measure of the times of peak of the 35-day cycles and converges well for the
BAT data. 
For the ASM data, the times of peak do not converge as well, with about half the times converging to better
than 0.1 day, and the other half showing a scatter of $\sim$1 day, as shown in Figure~\ref{fig:ASMtemp234}.

The average 35-day cycle length from BAT data is 34.79 d, with standard deviation, SD, of 1.1 d.
The ASM average cycle length is 34.55 d, with SD of 3.4 d.
The ASM SD is expected to be larger because of the larger timing errors from the CC analysis. 
The larger timing errors are confirmed by an analysis of the time period where Her X-1 was observed
with both BAT and ASM.

The BAT analysis gives the best measurement to date for orbital phase of turn-on.
The analysis revealed a distribution of orbital phases of peak of 35-day cycle 
which is well fit by the sum of a uniform and a Gaussian distribution.
 76\% of the cycles are in the uniform distribution, and 24\% of cycles are  in the Gaussian,
which has a maximum at orbital phase 0.51.
The larger errors in timing for the ASM data means that we do not detect the peak in the ASM analysis.
The previous result of no maximum from ASM data analysis \citep{2010ApJ...713..318L} is explained 
by timing errors caused by noise in the ASM data.
The maximum in orbital phase of peak corresponds to a maximum in orbital phase for turn-on of 0.85. 
This is different than the often reported peaks near 0.2 and 0.7 orbital phase reported in less complete studies
(e.g. \citealt{1999ApJ...510..974S} and references therein).

We present 35-day average lightcurves for BAT, in the 15-50 keV band, and for ASM, in the
2-12 keV band in Figure~\ref{fig:BATtemp}. 
These include $\sim14.5$ years of observation for each instrument and constitute
the best measures yet of the average 35-day cycle lightcurve.

The MH state and Short High state are more variable (higher SD) 
in the low energy band than in the high energy band. 
This is expected in the precessing accretion disk model for the 35-day cycle 
(\citealt{2002MNRAS.334..847L} and reference therein), and in the origin of dips from the
accretion stream (\citealt{2012MNRAS.425....8I} and references therein).
The reason is that absorption by cold matter is one of the main contributors to the shape and variability 
of the 35-day cycle.

The low (ASM) and high energy (BAT) bands of the average 35-day lightcurve exhibit 3 dips during the MH
state. These can be attributed to the increased dipping behaviour seen during MH as 35-day phase increases.
Consistent with this, the cycle to cycle SD is highest in the ASM band during these 3 dip periods.
As shown by \citet{2011ApJ...736...74L}, the system spends most of its time in dips during Short High,
thus the average of the dips with different timing does not show up as a regular feature during Short High 
and the SD is not as high in the ASM band as for Main High state.
The ASM and BAT lightcurves both have high SD at start of turn-on to MH.
The SD is high for both because the beginning of turn-on is variable in time and has such a high
column density ($>10^{24}$ cm$^{-2}$) that both energy bands are affected.

The observer's line of sight is close to the inner ionized edge of the accretion disk throughout Short High
(\citealt{2000ApJ...539..392S} and \citealt{2002MNRAS.334..847L}). The resulting scattering affects the
high energy band as much as the low energy band, causing the SD in the BAT energy band 
to be maximum for Short High.

In conclusion, the RXTE/ASM and Swift/BAT monitoring observations of Her X-1 have given an unprecedented
longterm view of the X-ray emission from this system. 
The properties of the 35-day cycle are now better measured than previously possible. 
Future work includes creating a complex enough model of the Her X-1 system, with enough 
physics for the accretion disk, to be able to explain the observed properties of the 35-day cycle including
the average light curves in the low and high energy bands.

\acknowledgments
This project is undertaken with the financial support of the Natural Sciences and Engineering Research Council of Canada.

\bibliography{HerX135dcycle}{}
\bibliographystyle{aasjournal}

\clearpage

\begin{table}
    \begin{center}
    \caption{BAT cycles. \label{tab:BATcycles}}
        \begin{tabular}{llllll}
    Cycle & Peak time (MJD) & Cycle length (d) & Orbital phase & Label$^1$ \\ \hline
       0 &         53437.82176 &    37.181 &   0.08430 &       0\\
       1 &         53475.00306 &    31.212 &   0.95347 &       0\\
       2 &         53506.21472 &    37.175 &   0.31144 &       0\\
       3 &         53543.38972 &    32.088 &   0.17691 &       0\\
       3 &         53543.38972 &    36.650 &   0.17691 &       0\\
       4 &         53575.47722 &    39.500 &   0.05002 &       2\\
       4 &         53580.03935 &    34.938 &   0.73336 &       2\\
       5 &         53614.97704 &    32.547 &   0.28289 &       0\\
       6 &         53647.52417 &    35.163 &   0.42635 &       0\\
       7 &         53682.68704 &    35.972 &   0.10832 &       0\\
       8 &         53718.65954 &    34.539 &   0.26650 &       0\\
       9 &         53753.19880 &    34.027 &   0.58169 &       0\\
    \end{tabular}
         \end{center}
    \tablecomments{$^1$ Label 0 stands for cycles detected by algorithm; 1 with detected orbital phase for weak cycles (relative low CC peak value and added manually); 1 with orbital phase assigned ``$-1.00000$" for calculated peak times (cycle lengths averaged with previous and later peak times); 2 for a range of time due to missing data in between (cycle number duplicated for the cycle and the one before it to show range of cycle lengths).
    Cycles with orbital phases marked ``$-1.00000$" are not identified by the cross-correlation algorithm. Their MJD time and cycle lengths are calculated from the MJD time of cycles right before and after. 
    The complete table with all 152 cycles is available in machine readable form.
 }
\end{table}

\begin{table}
    \begin{center}
    \caption{ASM cycles$^0$. \label{tab:ASMcycles}}
        \begin{tabular}{llllll}
    Cycle & Peak time (MJD) & Cycle length (d) & Orbital phase & Label$^1$ \\ \hline
       2 &         50146.55311 &    38.628 &   0.23719 &    1\\
       3 &         50185.18145 &    35.844 &   0.95747 &    0\\
       4 &         50221.02589 &    36.044 &   0.04033 &    0\\
       5 &         50257.06941 &    33.454 &   0.24028 &    0\\
       6 &         50290.52330 &    33.940 &   0.91708 &    0\\
       7 &         50324.46311 &    35.491 &   0.87968 &    0\\
       8 &         50359.95404 &    36.229 &   0.75461 &    0\\
       9 &         50396.18293 &    33.875 &   0.06360 &    0\\
      10 &         50430.05811 &    35.014 &   0.98819 &    0\\
      11 &         50465.07163 &    36.168 &   0.58232 &    0\\
    \end{tabular}
     \end{center}
       \tablecomments{$^0$ Numbering of cycle starts from 2 in order to better compare with Table 1 of Leahy \& Igna (2010). For the same reason cycle \# 50 is skipped. We use 17 cycles, each 35.744 days long, during the anomalous low state from MJD51223 to MJD51831, while Leahy \& Igna (2010) have 14 cycles of 33.4-day lengths from MJD51222 to MJD51757. The cycles would have lengths of 33.759 days if there are 18 cycles during this ALS time period. 
        The complete table with all 160 cycles is available in machine readable form.
 }
\end{table}

\begin{table}
    \centering
    \caption{BAT \& ASM Cycle lengths \label{tab:BAT-ASMcyclen}}
    \begin{tabular}{lllll}
 Analysis Period &  Data/Cycles & Average (d) & SD (d) \\  \hline 
All data  & BAT/CC & 34.787 & 1.1165 \\
All data &   ASM/CC & 34.553 & 3.4246 \\ 
\hline 
ASM/BAT overlap &   BAT/CC & 34.755 & 1.1764 \\
ASM/BAT overlap  &  ASM/CC & 34.983 & 1.0520 \\ \hline
    \end{tabular}
\end{table}

\begin{table}
    \begin{center}
\caption{ASM \& BAT Comparison for Overlap Period \label{tab:BATASMcycles}}
        \begin{tabular}{llllll}
\colhead{Cycle No.} & \colhead{Cycle No.}  & \colhead{ASM-BAT peak}  & \colhead{ASM-BAT orbital} \\
\colhead{for ASM} & \colhead{for BAT} & \colhead{time difference (d)} & \colhead{phase difference}  \\ 
     97  &       0  &             1.36578 &   0.80335\\
      98  &       1  &             1.01504 &  -0.40295\\
     100  &       3  &            -0.70143 &   0.58745\\
     101  &       4  &            -0.78496 &  -0.46167\\
     102  &       5  &            -1.23857 &   0.27152\\
     104  &       7  &             1.28828 &   0.75777\\
     105  &       8  &            -0.00792 &  -0.00463\\
     106  &       9  &            -0.55052 &  -0.32378\\
     107  &      10  &             0.25653 &   0.15092\\
     108  &      11  &             0.56874 &  -0.66546\\
    \end{tabular}
         \end{center}
     \tablecomments{
        The complete table with all 59 cycles is available in machine readable form.
 }
\end{table}

\end{document}